\documentclass[12pt]{iopart}

\usepackage{graphicx,bm}
\usepackage{txfonts}
\usepackage{color}
\usepackage{hyperref}

\newcommand{\rp}[1]{(\ref{#1})}

\newcommand{\abs}[1]{\left|{#1}\right|}

\newcommand{\rabs}[1]{\left.{#1}\right|}

\newcommand{\Br}[1]{\left\langle #1\right |}
\newcommand{\ke}[1]{|#1\rangle}
\newcommand{\Ke}[1]{\left |#1\right \rangle}

\newcommand{\KB}[2]{\Ke{#1}\Br{#2}}

\newcommand{\wt}[0]{\widetilde}

\newcommand{\al}[1]{^{(#1)}}
\newcommand{\da}{^\dagger}

\newcommand{\ppt}[1]{\left( #1 \right)}
\newcommand{\pq}[1]{\left[ #1 \right]}
\newcommand{\pg}[1]{\left\{ #1 \right\}}

\newcommand{\lpq}[1]{\left[ #1 \right.}
\newcommand{\lpg}[1]{\left\{ #1 \right.}

\newcommand{\rpq}[1]{\left. #1 \right]}
\newcommand{\rpg}[1]{\left. #1 \right\}}
\newcommand{\ee}{{\rm e}}
\newcommand{\ii}{{\rm i}}

\usepackage{bbold}

\newcommand{\nn}{{\nonumber}}

\newcommand{\mmat}[2]{
                      \begin{array}{#1}
                       #2
                       \end{array}  }

\newcommand{\ovl}{\overline}

\newcommand{\DD}{{\cal D}}

\newcommand{\LL}{{\cal L}}

\definecolor{green}{rgb}{0,0.6,0}

\usepackage[normalem]{ulem}

\newcommand{\stkout}[1]{\ifmmode\textrm{\sout{\ensuremath{#1}}}\else\sout{#1}\fi}

\begin{document}

\title[Dissipative stabilization of entangled qubit pairs in quantum arrays]
{Dissipative stabilization of entangled qubit pairs in quantum arrays with a single localized dissipative channel}

\author{Jacopo Angeletti$^{1,2,3}$, Stefano Zippilli$^1$, David Vitali$^{1,3,4}$}
\address{$^1$ School of Science and Technology, Physics Division, University of Camerino, I-62032 Camerino (MC), Italy\\
$^2$ Department of Physics, University of Naples ``Federico II", I-80126 Napoli, Italy\\
$^3$ INFN, Sezione di Perugia, I-06123 Perugia, Italy\\
$^4$ CNR-INO, I-50125 Firenze, Italy}

\vspace{10pt}
\begin{indented}
\item[]\today
\end{indented}

\begin{abstract}
We study the dissipative stabilization of entangled states in arrays of quantum systems. Specifically, we are interested in the states of qubits (spin-1/2) which may or may not interact with one or more cavities (bosonic modes). In all cases only one element, either a cavity or a qubit, is lossy and irreversibly coupled to a reservoir. When the lossy element is a cavity, we consider a squeezed reservoir and only interactions which conserve the number of cavity excitations. Instead, when the lossy element is a qubit, pure decay and a properly selected structure of $XY$-interactions are taken into account. We show that in all cases, in the steady state, many pairs of distant, non-directly interacting qubits, which cover the whole array, can get entangled in a stationary way, by means of the interplay of dissipation and local interactions.
\end{abstract}

\section{Introduction}
A central, necessary ingredient of quantum technologies such as quantum computation, simulation, and communication is the ability to control and distribute entangled resources over large arrays of quantum systems. An attractive strategy makes use of controlled dissipative processes to steer and protect arrays of quantum systems into entangled states~\cite{plenio1999,kraus2008,diehl2008,verstraete2009,weimer2010,barreiro2011,
cho2011a,koga2012,morigi2015a,reiter2016}. In particular, it has been shown that, in order to drive the whole system into non-trivial and potentially useful multipartite entangled states, it is sufficient to control the dissipative dynamics of one or two localized elements in a quantum array
~\cite{zippilli2013_zippilli2013a,zippilli2014,ma2019,
wendenbaum2015,pocklington2022,zippilli2015,asjad2016a,ma2017,
yanay2018,yanay2020b,yanay2020a,zippilli2021,ma2021}.
This has been proved for arrays of both bosonic (by using a single localized squeezed reservoir)~\cite{zippilli2015,asjad2016a,ma2017,yanay2018,yanay2020b,yanay2020a,zippilli2021} and fermionic (via a correlated reservoir for two fermions)~\cite{pocklington2022} modes. It has also been shown that many entangled pairs can be realized by modulating the coupling between a central cavity and two qubits (spin-1/2) in a qubit chain~\cite{ma2021} and by designing a correlated reservoir of two elements in an array of qubits~\cite{zippilli2013_zippilli2013a,wendenbaum2015,pocklington2022} and cavities (bosonic modes)~\cite{zippilli2013_zippilli2013a,zippilli2014,ma2019}.

Here we present that, also in the case of qubits, it is sufficient to control the local environment of a single element of an array to generate, in the steady state, many entangled qubits pairs. We demonstrate this both for arrays of cavities and qubits, and for arrays of only qubits. In the first case a single cavity is coupled to a squeezed reservoir (see Ref.~\cite{Clark_2017} for an experimental implementation in the microwave domain and Refs.~%
\cite{virgocollaboration2019,tse2019}
for its use in the optical domain to improve the performance of gravitational wave interferometers) and all the interactions conserve the number of excitations. In the second, instead, a single qubit can decay, and the qubits are coupled according to a specific geometry of $XY$-interactions. Ideally, we assume that only one element of the arrays is lossy. We also analyze how these dynamics are sensitive to additional noise affecting the qubits.

We note that similar states have also been found in the ground states of certain spin Hamiltonians~\cite{vitagliano2010,ramirez2014,langlett2022,zippilli2013b}, i.e., the so-called concentric singlet phase~\cite{vitagliano2010}
and rainbow states~\cite{ramirez2014,pocklington2022}.
They can also be generated in a spin chain following specific dynamics~\cite{difranco2008,alkurtass2014,pitsios2017},
and in this context they have been labeled nested entangled states (or matryoshka states)~\cite{difranco2008}. Moreover, analogous states are also known as thermofield double states in the high-energy community~\cite{cottrell2019,brown2021}.
Differently from all these examples, here we display that these states can be found as the \emph{unique} pure steady state of a dissipative dynamics.

Finally, we mention the related results reported in Refs.~\cite{dutta2020,dutta2021} where however the entangled steady states are not unique, so that they are achieved only if the system is prepared in peculiar initial states. 

The outline of the paper is as follows. In Sec.~\ref{Sec.c+q} we describe in detail the four models of arrays involving both qubits and cavities, and describe also the main result, that is, the possibility to generate in a robust way a stationary state of many entangled qubit pairs. In Sec.~\ref{Sec.q} we describe how a similar dissipative generation of entangled qubit pairs can be obtained with effective models involving only qubits. In Sec.~\ref{num} we verify 
our results through the numerical solution of the dynamics of all the models presented in the previous sections, while Sec.~\ref{concl} is for concluding remarks.

\section{Models with qubits and cavities}
\label{Sec.c+q}
In this section we analyze the models which involve both qubits and cavities.
We identify four models which differ in geometries and composition of the arrays as specified below. In all cases there is a central cavity which is coupled to a squeezed reservoir, and the corresponding stationary state is pure and factorized between the state of the qubits $\ke{\psi}$, and that of the cavity/ies  $\ke{\varphi}_c$. In particular, without exception, each qubit gets entangled to another one, such that they form many entangled pairs. In details, by using a proper labeling, each qubit with label $j$ gets entangled with the $-j$th, and the steady state of the qubits can be expressed as
\begin{eqnarray}\label{psiq}
    \ke{\psi}=\bigotimes_{j=1}^N 
    \ppt{
    \sqrt{\frac{\ovl n+1}{2\,\ovl n+1}}\ \ke{-}_j\ke{-}_{-j}-
    \chi_j\
    \sqrt{\frac{\ovl n}{2\,\ovl n+1}}\ \ke{+}_j\ke{+}_{-j}
    }\ ,
    \nn\\
\end{eqnarray} 
where the variable $j$ runs over all the entangled pairs, $N$ is the number of pairs (such that the number of qubits is $2\,N$), $\ke{\pm}_j$ is the eigenvalue of the Pauli operator $\sigma_j\al{z}$, for the qubit $j$, with eigenvalue $\pm 1$, and $\ovl n$ is the number of excitations of the squeezed reservoir. Moreover, $\chi_j$ is a phase factor which depends on the specific model [as specified in Eq.~\rp{chij}]. We finally note that, by increasing $\ovl n$, the state of each pair in Eq.~\rp{psiq} tends to a Bell, maximally-entangled state, which is a central resource in many quantum information protocols~\cite{horodecki2009}.
\begin{figure}[t!]
    \centering
    \includegraphics[width=12cm]{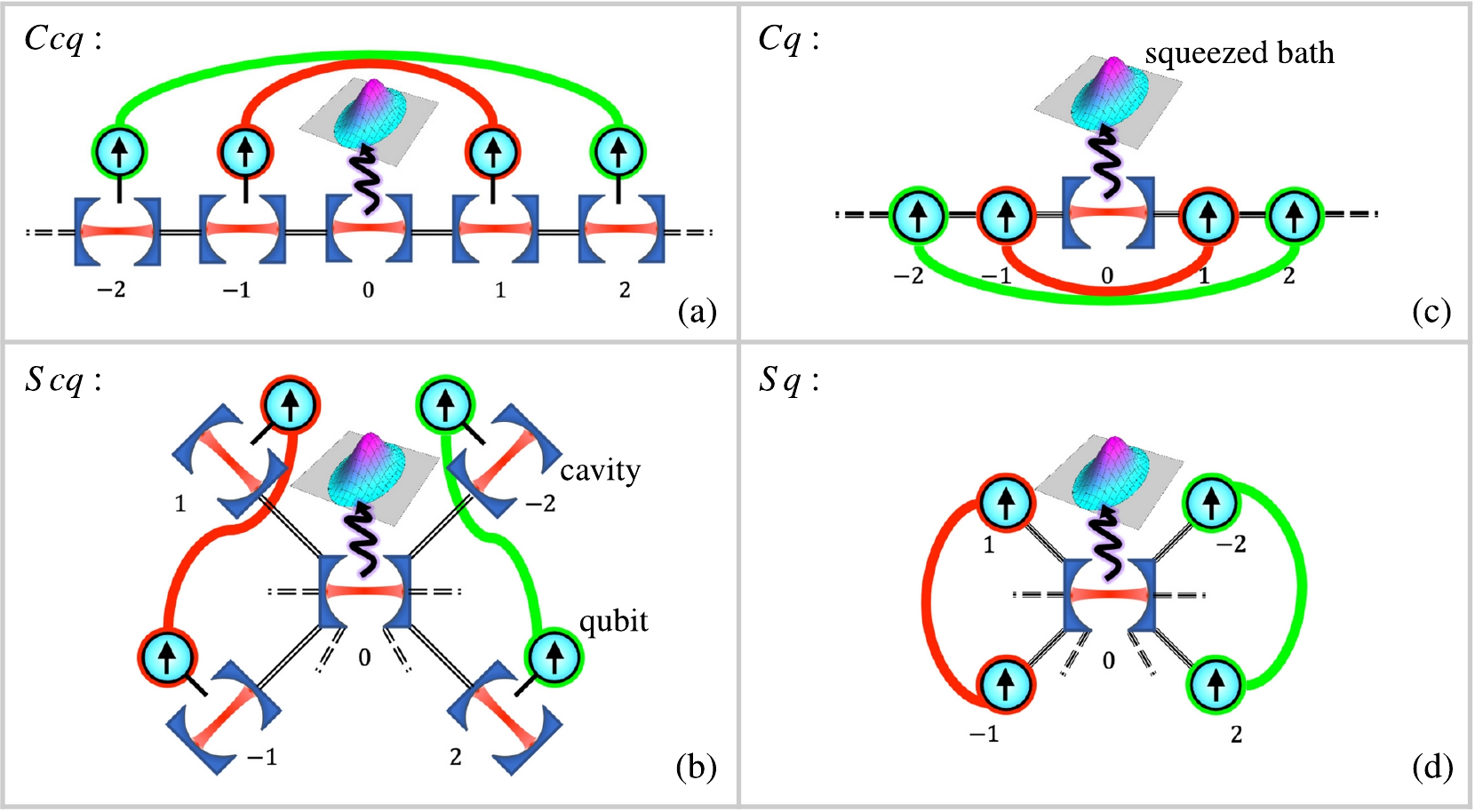}
    \caption{(a) Chain of cavities and qubits ``$Ccq$'' where each cavity but the central one interacts with a qubit. The central cavity is locally coupled to a squeezed bath. Regardless of the initial conditions, steady-state entangled pairs of qubits (pairs indicated by the red and green thick lines) arise. (b) Star geometry with cavities and qubits ``$Scq$''. (c) Chain of qubits with a central cavity ``$Cq$''. (d) Star of qubits with a central cavity ``$Sq$''.
    }
	\label{fig1}
\end{figure}

\subsection{The four models}

\paragraph{Chain of cavities and qubits.}
The first model is an extension of the chain of cavities studied in Ref.~\cite{zippilli2015} where, here, each cavity but the central one interacts also with a qubit [see Fig.~\ref{fig1} (a)]. In the following we indicate this model with the symbol ``$Ccq$'', where the upper case ``$C$'' stands for chain and thus addresses the geometry, while ``$cq$'' the composition of the array, i.e., of cavities and qubits. 
 
\paragraph{Star of cavities and qubits.}
Similarly, the second model is the extension of a star--like bosonic array analogous to that discussed in Ref.~\cite{asjad2016a} where, here, each of the external modes interacts with a qubit [see Fig.~\ref{fig1} (b)]. For this model we use the symbol ``$Scq$'' where now the upper case ``$S$" indicates the star geometry. 

\paragraph{Chain of qubits with a central cavity.}
Then, we consider a chain of qubits, where the central element is in fact a cavity (or, in other terms, it is made of two chains coupled on one end to a common cavity) [see Fig.~\ref{fig1} (c)]. In this case we use the symbol ``$Cq$'' for chain of qubits.  

\paragraph{Star of qubits with a central cavity.}
Finally we also study a star--like array where a central cavity is coupled to many qubits [see Fig.~\ref{fig1} (d)]. For this model we use the symbol ``$Sq$''.

\subsection{The master equation}
In every case, the system dynamics is described by a master equation of the form
\begin{eqnarray}\label{Meq}
    \dot\rho=-\frac{\ii}{\hbar}\pq{H_\xi,\rho}+\frac{\kappa}{2}\,\LL_c\,\rho\ ,
\end{eqnarray}
with $\xi\in\pg{Ccq,Scq,Cq,Sq}$. The Lindblad operator $\LL_c$ describes dissipation via the squeezed reservoir of the central cavity with photonic annihilation operator $b_0$ (see Fig.~\ref{fig1}), and it reads 
\begin{eqnarray}
    \LL_c\ \rho= 2\ \beta_0\ \rho\ \beta_0\da-\beta_0\da\ \beta_0\ \rho-\rho\ \beta_0\da\ \beta_0\ ,
\end{eqnarray}
where
\begin{eqnarray}
    \beta_0=\sqrt{\,\ovl n+1}\ b_0-\sqrt{\,\ovl n}\ b_0\da\ ,
\end{eqnarray}
is the squeezed annihilation operator. A squeezed reservoir can be realized by driving the system with a broadband squeezed field~\cite{zippilli2014,asjad2016a,asjad2016b}. In the microwave regime, a squeezed reservoir has been experimentally realized, as reported in Ref.~\cite{Clark_2017}. This technique has also been employed to enhance the performance of gravitational wave interferometers in the optical domain~\cite{virgocollaboration2019,tse2019}.

\subsection{The Hamiltonians of the four models}\label{Sec:H}
In all instances, we take into account only Hamiltonian interactions which conserve the number of excitations. Thus, for two interacting cavities with annihilation operators $b_j$ and $b_k$, we consider interaction Hamiltonians of the form $H_{c-c}\propto b_j\da\ b_k+h.c.$ (where $h.c.$ indicates the Hermitian conjugate). Moreover, the cavity-qubit interactions are described by Jaynes-Cummings Hamiltonians $H_{c-q}\propto b_j\da\,\sigma_j+h.c.$, with $\sigma_j=\pq{\sigma_j\al{x}-\ii\,\sigma_j\al{y}}/2$ the lowering operator for the qubit $j$. Finally, the interactions between two qubits are described by an $XX$ spin-1/2 Hamiltonian $H_{q-q}\propto\sigma_j\al{x}\,\sigma_k\al{x}+\sigma_j\al{y}\,\sigma_k\al{y}=2\,\sigma_j\da\ \sigma_k+h.c. $.

To be specific, the Hamiltonians corresponding to each of the four models are given by the expressions reported in Eqs.~\rp{HCcq}-\rp{HSq} below, which describe the system in a reference frame rotating at the frequency of the central cavity $\omega_0/2\pi$. 
In particular - on the one hand - in the models ``$Ccq$" and ``$Scq$" with many cavities (each interacting with a qubit), all the qubits are resonant with the central cavity, while the other cavities are detuned by a frequency $\Delta_{c,j}$ from $\omega_0$. On the other hand, in the models ``$Cq$" and ``$Sq$", composed of many qubits and a central cavity, the transition frequency of each qubit is detuned by $\Delta_{q,j}$ from $\omega_0$. Our analysis relies on the assumption that $\omega_0$ is the dominant parameter in the studied systems. Consequently, the frequencies of the cavities and qubits ($\omega_0+\Delta_{c,j}$ and $\omega_0+\Delta_{q,j}$, respectively) are orders of magnitude larger than the introduced coupling strengths, a common feature of quantum-optical systems. This key assumption justifies our adoption of the local master Eq.~(\ref{Meq})~\cite{LocalGlobal_MEQ}.

We also note that, in order to sustain the steady state of Eq.~\rp{psiq}, the Hamiltonians have to fulfill certain symmetry properties: they have to be symmetric in the interaction strengths and antisymmetric in the detunings as discussed below. This is analogous to the chiral symmetry identified in Ref.~\cite{pocklington2022} (see also Ref.~\cite{yanay2018}) and which is at the basis of the emergence of steady state entangled pairs (equal to the ones studied here) in a finite chain of qubits, when the two central qubits are coupled to a correlated reservoir. 

\

Let us now introduce the explicit formulas for the Hamiltonians. 

\paragraph
{Chain of cavities and qubits: ``$Ccq$" [Fig.~\ref{fig1} (a)]}

\begin{eqnarray}\label{HCcq}
    H_{Ccq}&=&
    \hbar\sum_{j=1}^N
    \lpg{
    \Delta_{c,j}\ppt{b_j\da\ b_j-b_{-j}\da\ b_{-j}} +
    \lpq{
    g_j\ppt{b_j\da\,\sigma_j+b_{-j}\da\,\sigma_{-j}}
    }}\nn\\&&\rpg{\rpq{
    +\eta_{c,j}\ppt{b_j\da\,b_{j-1}+b_{-j}\da\,b_{-j+1}}+h.c.
    }}\ ,
\end{eqnarray}
where $j=0$ indicates the central cavity, while the positive and negative values of $j$ indicate the elements respectively on the right and on the left of the central cavity, with the value of $\abs{j}$ measuring the distance from the central cavity. Here we see that cavities at the same distance on the right and on the left have opposite detuning, while all the interactions are symmetric.

In such a situation (and also in the following), both qubits and cavities form many entangled pairs. The dynamics of the cavities is the same as that discussed in Ref.~\cite{zippilli2015} (which specializes Ref.~\cite{zippilli2021}). Here we demonstrate how the entanglement of these bosonic modes is transferred to the qubits, similarly to Refs.~\cite{zippilli2013_zippilli2013a,
benatti2003_kraus2004_paternostro2004_adesso2010}.

\
\paragraph
{Star of cavities and qubits: ``$Scq$" [Fig.~\ref{fig1} (b)]}

\begin{eqnarray}\label{HScq}
    H_{Scq}&=&\hbar\sum_{j=1}^N
    \lpg{
    \Delta_{c,j}\ppt{b_j\da\ b_j-b_{-j}\da\ b_{-j}} +
    \lpq{
    g_j\ppt{b_j\da\,\sigma_j+b_{-j}\da\,\sigma_{-j}}
    }}\nn\\&&\rpg{\rpq{
    +\eta_{c,j}\ b_0\da\ppt{b_j+b_{-j}}+h.c.
    } }\ .
\end{eqnarray}
In this regard the index $j$ does not indicate the distance from the central cavity, but it is still used to label the elements which are entangled in the steady state. In other words, the elements (both cavities and qubits) with indices $j$ and $-j$ form entangled pairs. This model shares various features with the previous one. First, the cavities are entangled in the steady state and follow a dynamics similar to that discussed in Ref.~\cite{zippilli2021} (see also Ref.~\cite{asjad2016a}) and this induces entanglement in the qubits as in the previous case~\cite{zippilli2013_zippilli2013a,
benatti2003_kraus2004_paternostro2004_adesso2010}. Second, the cavities in each pair have opposite detuning, while their interaction coefficients are equal.

\
\paragraph
{Chain of qubits with a central cavity: ``$Cq$" [Fig.~\ref{fig1} (c)]}

\begin{eqnarray}\label{HCq}
    H_{Cq}&=&\hbar\sum_{j=1}^N
    \frac{\Delta_{q,j}}{2}\pq{\sigma_j\al{z}-\sigma_{-j}\al{z}} 
    +
    \hbar
    \lpq{
    \vphantom{\sum_{j=2}^N}
    g_1\ b_0\da \ppt{\sigma_1+\sigma_{-1}}
    }\nn\\&&\rpq{
    +\sum_{j=2}^{N}\ \eta_{q,j}\ppt{\sigma_j\da\,\sigma_{j-1}+ \sigma_{-j}\da\,\sigma_{-j+1} }+h.c.
    }\ ,
\end{eqnarray}
where, similar to the ``$Ccq$'' model, the positive and negative values of $j$ indicate, respectively, \textcolor{yellow}{}the qubits on the right and on the left of the central cavity. A qubit on the right chain has a detuning which is opposite to that of the corresponding qubit on the left chain, while the corresponding couplings are identical. As described earlier, pairs of qubits at the same distance on the right and on the left of the cavity are entangled in the stationary state.

\
\paragraph
{Star of qubits with a central cavity: ``$Sq$" [Fig.~\ref{fig1} (d)]}

\begin{eqnarray}\label{HSq}
    H_{Sq}&=&\hbar\sum_{j=1}^N
    \pg{\frac{\Delta_{q,j}}{2}\pq{\sigma_j\al{z}-\sigma_{-j}\al{z}} 
    +\pq{g_j\,b_0\da\ppt{\sigma_j+\sigma_{-j}}+h.c.} 
    }\ .
    \nn\\
\end{eqnarray}
As in the ``$Scq$" case, the index $j$ does not indicate the distance from the central cavity, but it is used to label the elements which are entangled in the steady state, that is, the qubits with indices $j$ and $-j$ form entangled pairs.

\subsection{The steady state}\label{Sec:stst}
Let us now analyze in detail the steady state of the models defined above. When $g_j=0$ (indicating no interaction with the qubits), the squeezed reservoir drives the central cavity towards a squeezed state. Correspondingly, as shown in Refs.~\cite{zippilli2015,asjad2016a,zippilli2021}, the other cavities (in the models ``$Ccq$'' and ``$Scq$'') approach a pure entangled state constituted of many two-mode squeezed states. It can be expressed as 
\begin{eqnarray}\label{psic}
    \ke{\,\varphi}_c=U_c\ke{0}_c,
\end{eqnarray}
where $\ke{0}_{c}$ is the vacuum and $U_c$ the unitary which generates the steady state. In detail, for the models of cavities and qubits (``$Ccq$'' and ``$Scq$''), $U_c$ is the product of a squeezing operator on the central cavity mode, and many two-mode squeezing operators for the modes with opposite indices
\begin{equation}\label{Uc1}
    U_c=\ee^{\frac{r}{2}\ppt{{b_0\da}^2-b_0^2}}\
    \bigotimes_{j=1}^N
    \ee^{
    \chi_j
    \,r\,\ppt{b_j\da\,b_{-j}\da-b_j\,b_{-j}}}\ ,
    \hspace{0.5cm} {\rm for}\ \ Ccq\ \ {\rm and}\ \ Scq\ ,
\end{equation}
where $\tanh(r)=\sqrt{\,\ovl n/\ppt{\,\ovl n+1}}$.
Instead, for the models ``$Cq$" and ``$Sq$", $U_c$ is the single mode squeezing operator
\begin{eqnarray}\label{Uc2}
    U_c=\ee^{\frac{r}{2}\ppt{{b_0\da}^2-b_0^2}}\ , \hspace{0.5cm} {\rm for}\ \ Cq\ \ {\rm and}\ \ Sq\ .
\end{eqnarray} 
In particular the 
factor $\chi_j$, which appears in Eqs.~\rp{psiq} and~\rp{Uc1}, is given by
\begin{equation}\label{chij}
	\chi_j=\left\{
 \mmat{llll}{
    &(-1)^j\ ,      & &{\rm for}\ \ Ccq\ ,\\
    &(-1)^{j+1}\ ,  & &{\rm for}\ \ Cq\ ,\\
    &-1\ ,          & &{\rm for}\ Scq\ ,\\
    &1\ ,           & &{\rm for}\ Sq}\ ,\\
	\right.
\end{equation}%
for $j\neq 0$. We also note that these operators [Eqs.~\rp{Uc1} and~\rp{Uc2}] realize the Bogoliubov transformation 
\begin{eqnarray}\label{provlab}
    U_c\da\ b_j\ U_c&=&\sqrt{\,\ovl n+1}\ b_j+
    \chi_j\
    \sqrt{\,\ovl n}\ b_{-j}\da\ ,
\end{eqnarray}
for all $j$, also for $j=0$ with $\chi_0 =1$. 

Now, one can check that, in general, the product state
\begin{eqnarray}\label{Psi}
    \ke{\Psi}=\ke{\,\varphi}_c\ \ke{\psi}\ ,
\end{eqnarray}
between the state of the qubits given by Eq.~\rp{psiq}, and that of the cavity/ies given by Eq.~\rp{psic}, with $U_c$ and $\chi_j$ defined in Eqs.~\rp{Uc1}-\rp{chij}, \emph{is a steady state for the four models}. In other words, one finds that $-\frac{\ii}{\hbar}\pq{H_\xi,\KB{\Psi}{\Psi}}+\frac{\kappa}{2}\,\LL_c\,\KB{\Psi}{\Psi}=0$. 

This is the result of the destructive interference which takes place when these systems are endowed with the specific symmetries described in Sec.~\ref{Sec:H}. To be specific, by dividing the Hamiltonians as the sum of the term which involves only the cavity operators $H_{c,\xi}$ (this is non-zero only for the models ``$Ccq$'' and ``$Scq$''), the one for the qubits alone $H_{q,\xi}$ (this is non-zero only for the models ``$Cq$'' and ``$Sq$'') and the terms which describe the interactions between cavity/ies and qubits $H_{c-q,\xi}$, 
such that 
\begin{eqnarray}
    H_\xi=H_{c,\xi}+H_{q,\xi}+H_{c-q,\xi}\ ,
\end{eqnarray}
we find that 
\begin{eqnarray}
    -\frac{\ii}{\hbar}\pq{H_{c,\xi},\Ke{\,\varphi}_{c\!}\Br{\varphi}}+\frac{\kappa}{2}\,\LL_c\,\Ke{\,\varphi}_{c\!}\Br{\varphi}=0\ ,
\end{eqnarray}
as demonstrated in Refs.~\cite{zippilli2015,yanay2018,zippilli2021}, and
\begin{eqnarray}
    H_{q,\xi}\,\ke{\psi}&=&0\ ,
    \\
    H_{c-q,\xi}\,\ke{\Psi}&=&0\ ,
\label{HcqxiPsi}
\end{eqnarray}
because of the destructive quantum interference between transitions which involve the qubit states in the quantum superposition of~Eq.~\rp{psiq}.

\subsection{Dynamics in the squeezed representation}

\begin{figure*}[t!]
    \centering
    \includegraphics[width=16cm]{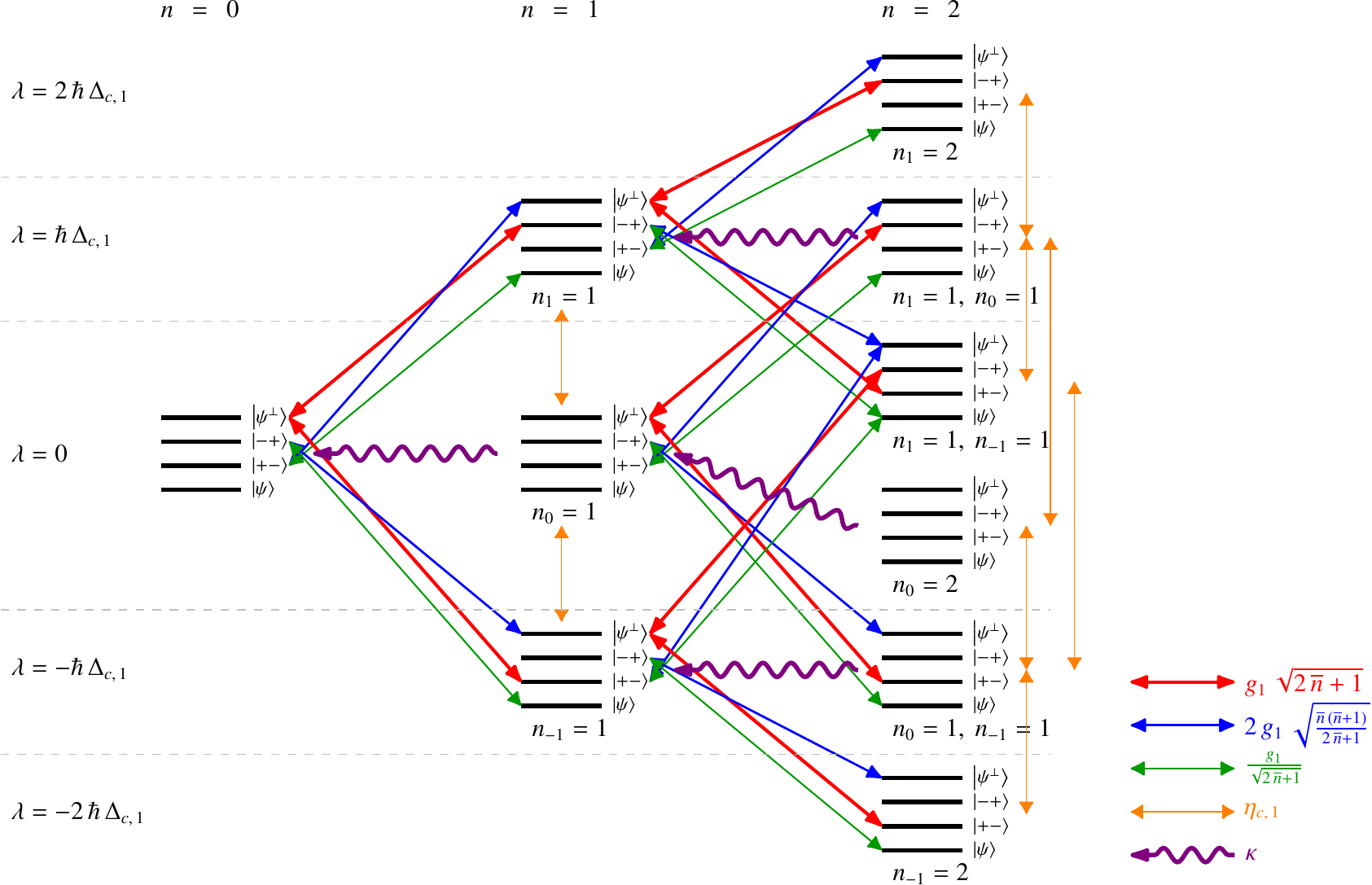}
    \caption{
    Low-excitations eigenlevels (black horizontal lines) corresponding to the Hamiltonian $\rabs{\wt H_{Ccq}}_{\eta_{c,j}=0,g_j=0}$ (in the squeezed representation) without interaction terms [see Eqs.~\rp{HCcq}, \rp{tildeHxi} and~\rp{tildeHcqxi}] and corresponding matrix elements of the interaction terms (red, blue, green, and orange arrows; each color marks a different coupling strength as shown in the right-bottom corner of the figure), for the model ``$Ccq$'' with $N=1$ (which is equal to the model ``$Scq$'' with $N=1$). The purple wavy arrow illustrates the transfer of population due to the decay of the central cavity; $\ke{\psi}=\ppt{\sqrt{\,\ovl n+1}\ \ke{--}+\sqrt{\,\ovl n}\ \ke{++}}/\sqrt{2\,\ovl n+1}$ is the qubit state of Eq.~\rp{psiq} and $\ke{\psi^\perp}=\ppt{\sqrt{\,\ovl n}\ \ke{--}-\sqrt{\,\ovl n+1}\ \ke{++}}/\sqrt{2\,\ovl n+1}$ is its orthogonal state. The parameters $n_j$ and $n\equiv\sum_j\,n_j$ indicate the Fock states $\ke{n_{-1}\ n_0\ n_1}_c$ of the cavities in the squeezed representation (i.e., the squeezed Fock states $U_c\ke{n_{-1}\ n_0\ n_1}_c$ in the original representation). The values of $\lambda$ tag the eigenvalues of $\rabs{\wt H_{Ccq}}_{\eta_{c,j}=0,g_j=0}$ corresponding to the various group of levels.  The horizontal dashed gray lines divide the groups of levels associated to each eigenvalue. Levels corresponding to the same eigenvalue (in the same group) are reported at different vertical positions in order to provide a better visualization.
    The state $\ke{\psi}$ with zero excitations, $n=0$, is the steady state of the system dynamics, where the population accumulates. In fact, being the only state decoupled from all the other levels, it does not lose population but it is populated by the cavity decay. 
    }
    \label{fig2}
\end{figure*}
In order to verify Eq.~\rp{HcqxiPsi} and gain insight into the steady state dynamics, it is useful to analyze the system in the representation in which the cavity steady state is the vacuum. Namely, we consider the master equation for the transformed density matrix $\wt\rho=U_c\da\,\rho\,U_c$, which is given by 
\begin{eqnarray}\label{tildeMeq}
    \dot{\wt\rho}=-\frac{\ii}{\hbar}\pq{\wt H_\xi,\wt \rho}+\frac{\kappa}{2}\,\wt\LL_c\,\wt\rho\ ,
\end{eqnarray}
with the Lindblad term which describes dissipation of the zeroth mode being
\begin{eqnarray}
    \wt\LL_c\ \wt\rho=2\ b_0\ \wt\rho\ b_0\da-b_0\da\ b_0\ \wt\rho-\wt\rho\ b_0\da\ b_0\ .
\end{eqnarray}
The transformed Hamiltonian in Eq.~\rp{tildeMeq} can be written as
\begin{eqnarray}\label{tildeHxi}
    \wt H_\xi&=&U_c\da\ H_\xi\ U_c
    \nn\\
    &=& H_{c,\xi}+H_{q,\xi}+\wt H_{c-q,\xi}\ ,
\end{eqnarray}
where the Jaynes-Cummings interaction term $\wt H_{c-q,\xi}=U_c\da\ H_{c-q,\xi}\ U_c$ may be expressed as 
\begin{eqnarray}\label{tildeHcqxi}
    \wt H_{c-q,\xi}
    &=&
    \hbar\sum_{j=1}^N
    \ g_j\ppt{b_j\da\ \tau_j+b_{-j}\da\ \tau_{-j}+h.c.}\ ,
    \hspace{0.5cm}{\rm for}\ \xi\in\pg{Ccq,Scq}\ ,
    \nn\\
    \wt H_{c-q,Cq}
    &=&
    \hbar\ g_1\ b_0\da\ppt{\tau_1+\tau_{-1}}+h.c.\ ,
    \nn\\
    \wt H_{c-q,Sq}
    &=&
    \hbar\sum_{j=1}^N\ g_j\pq{b_0\da\ppt{\tau_j+\tau_{-j}}+h.c.}\ ,
\end{eqnarray}
with $\tau_j$ given by the collective qubit operator
\begin{eqnarray}
    \tau_j=\sqrt{\,\ovl n+1}\ \sigma_j+
    \chi_j\
    \sqrt{\,\ovl n}\ \sigma_{-j}\da\ .
\end{eqnarray}
Now it is easy to verify that the transformed state
\begin{eqnarray}\label{tildePsi}
    \wt{\ke{\Psi}}=\wt{\ke{\,\varphi}}_c\ \ke{\psi}=\ke{0}_c\ \ke{\psi}\ ,
\end{eqnarray}
(with $\wt{\ke{...}}=U_c\da\,\ke{...}$) fulfills the relation $\wt H_{c-q,\xi}\,\wt{\ke{\Psi}}=0$ [which is equivalent to Eq.~\rp{HcqxiPsi}]. In fact,
on the one hand, $b_j\,\wt{\ke{\Psi}}=0$ because the cavity(-ies) is (are) in the vacuum (in this representation) and, on the other hand,
\begin{eqnarray}
    \tau_j\,\ke{\psi}=0\ ,\hspace{1cm}\forall\ j\ .
\end{eqnarray}
In other terms, in this representation, all the cavities dissipate (through the central cavity) and approach the vacuum. Correspondingly the qubits population accumulates in the entangled state of Eq.~\rp{psiq} in a way similar to optical pumping. In fact, the state of Eq.~\rp{tildePsi} is the only one which does not decay and it is decoupled from all the others, and in the meanwhile it is populated by the decay of the central cavity.

It is instructive to visualize this with a simple example, i.e., when $N=1$, for the models ``$Ccq$" and ``$Scq$" which are equal. In Fig.~\ref{fig2} we report the eigenstates of the system Hamiltonian without interactions and use arrows to connect levels which are coupled by the interaction terms. The number of cavity excitations (in the squeezed representation) increases from left to right, and the dissipation of the central cavity is responsible for an irreversible transfer of population from the levels on the right to those on the left. The figure shows that the qubit state of Eq.~\rp{tildePsi} [i.e., the state of Eq.~\rp{psiq} with zero cavity excitations] is the only state which is decoupled from the other levels, and at the same time it does not decay and is populated by the cavity decay. As a consequence this state is stable in the steady state. Similar considerations hold also for the other models.

\section{Models with only qubits}\label{Sec.q}
\begin{figure}[t!]
    \centering
    \includegraphics[width=12cm]{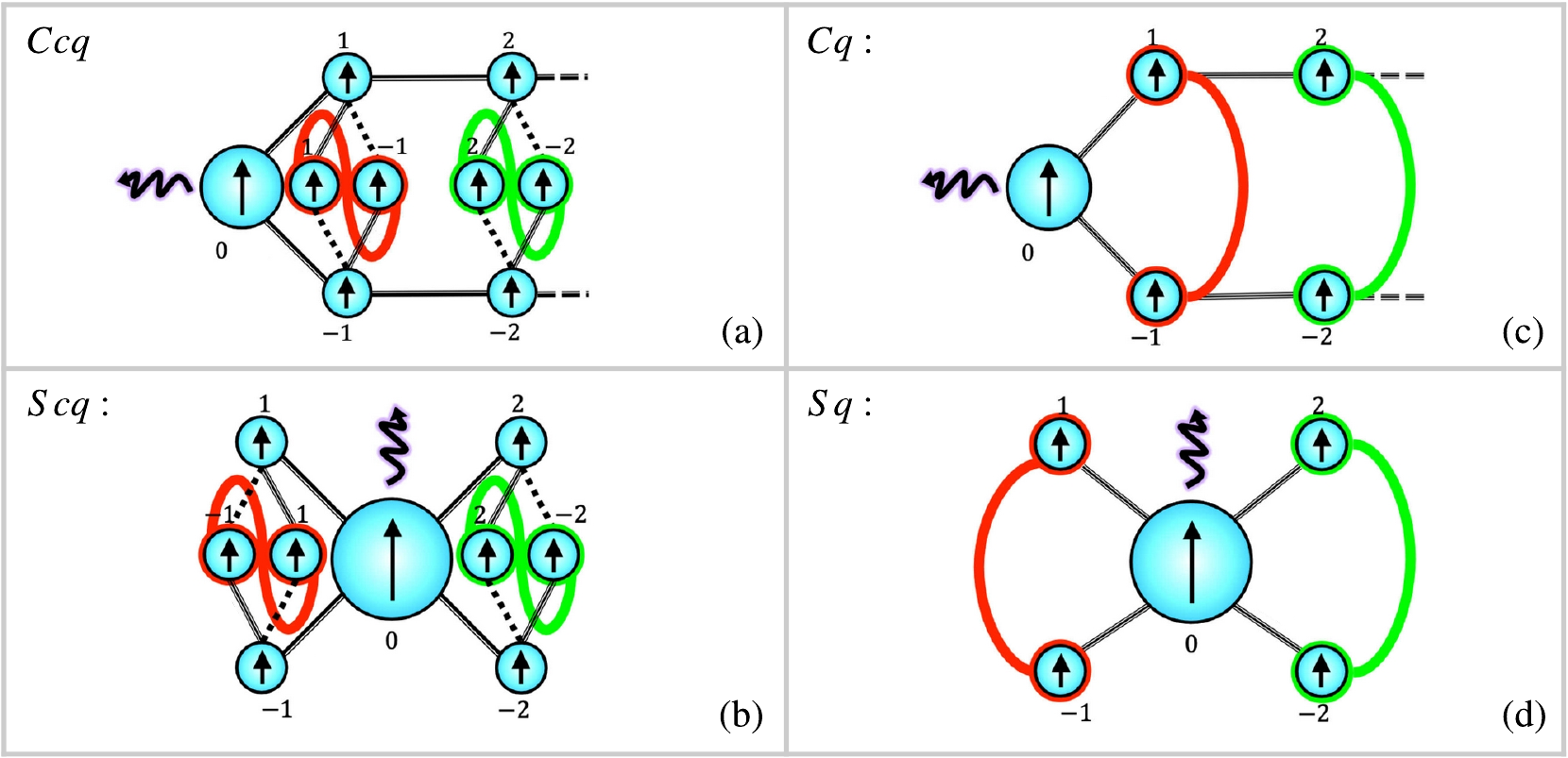}
    \caption{
    Sketch of the arrays of qubits analogous of the models of Fig.~\rp{fig1} but with the cavities replaced by additional qubits. In each model the central qubit is lossy (described by the wavy line) and the qubits which are entangled in the steady state are indicated by the red and green thick lines.
    In (a) and (b) the solid black lines connecting two qubits indicate $XX$- (isotropic $XY$-) interactions $\propto\sigma_j\al{x}\sigma_k\al{x}+\sigma_j\al{y}\sigma_k\al{y}$, while the dotted black lines indicate interactions of the form $\propto\sigma_j\al{x}\sigma_k\al{x}-\sigma_j\al{y}\sigma_k\al{y}$. In (c) and (d) the interactions between the central spin and the neighboring ones is of $XY$-type with anisotropic couplings $\propto g\al{x}\,\sigma_j\al{x}\sigma_k\al{x}+g\al{y}\,\sigma_j\al{y}\sigma_k\al{y}$, while the black lines in (c) between the other spins indicate $XX$-interactions $\propto\sigma_j\al{x}\sigma_k\al{x}+\sigma_j\al{y}\sigma_k\al{y}$.
    }
    \label{fig3}
\end{figure}

A similar qubits dynamics (in the squeezed representation) is observable also if we modify our models by replacing all the cavity modes with fresh new qubits. Namely, we may consider the master Eq.~\rp{tildeMeq} and replace all the $b_j$ and $b_j\da$ with other lowering and rising qubits operators $\sigma_{c,j}$ and $\sigma_{c,j}\da$ (where the label $_c$ indicates that these are the new qubits in place of the cavities of Sec.~\ref{Sec.c+q} and Fig.~\ref{fig1}). In this way we find new models which consist of only qubits, with a single central lossy one. Moreover, the qubits interact with a peculiar structure (different for each model) of $XY$-interactions (see Fig.~\ref{fig3}).

Let us explicitly write down the equations for these models.  The master equation has the form of Eq.~\rp{tildeMeq}
\begin{eqnarray}
    \dot{\varrho}=-\frac{\ii}{\hbar}\pq{H_\xi^\circ,\varrho}+\frac{\kappa}{2}\,\LL_q\,\varrho\ ,
\end{eqnarray}
where also in this case we use the labels $\xi\in\pg{Ccq,Scq,Cq,Sq}$ to highlight the relation with the models of Fig.~\ref{fig1},
but where now the Hamiltonian and the Lindblad operator include only qubits operators, following the substitution $b_j\to\sigma_{c,j}$. In other words
\begin{eqnarray}
    \LL_q&=&\rabs{\wt \LL_c}_{b_j\to\sigma_{c,j}}\ ,
    \nn\\
    H_\xi^\circ&=&\rabs{\wt H_\xi}_{b_j\to\sigma_{c,j}}\ ,
\end{eqnarray}
namely
\begin{eqnarray}
    \LL_q\ \varrho=2\ \sigma_{c,0}\ \varrho\ \sigma_{c,0}\da-\sigma_{c,0}\da\ \sigma_{c,0}\ \varrho-\varrho\ \sigma_{c,0}\da\ \sigma_{c,0}\ ,
\end{eqnarray}
and 
\begin{eqnarray}
    H_\xi^\circ&=&H_{c,\xi}^\circ+H_{q,\xi}+H_{c-q,\xi}^\circ\ ,
\end{eqnarray}
where, as in Sec.~\ref{Sec.c+q}, $H_{q,\xi}$ is non-zero only for $\xi\in\pg{Cq,Sq}$ and $H_{c,\xi}^\circ$ is non-zero only for $\xi\in\pg{Ccq,Scq}$, with
\begin{eqnarray}
    H_{c,Ccq}^\circ&=&\hbar\sum_{j=1}^N\sum_{\zeta=\pm}
    \pq{
    \zeta\Delta_{c,j}\
    \sigma_{c,\zeta j}\al{z}
    +\frac{\eta_{c,j}}{2}\ppt{
    \sigma_{c,\zeta j}\al{x}\,\sigma_{c,\zeta(j-1)}\al{x}+\sigma_{c,\zeta j}\al{y}\,\sigma_{c,\zeta(j-1)}\al{y}
    }}\ ,
    \nn\\
    H_{c,Scq}^\circ&=&\hbar\sum_{j=1}^N\sum_{\zeta=\pm}
    \pq{
    \zeta\Delta_{c,j}\
    \sigma_{c,\zeta j}\al{z}
    +\frac{\eta_{c,j}}{2}\ppt{
    \sigma_{c,0}\al{x}\,\sigma_{c,\zeta j}\al{x}+\sigma_{c,0}\al{y}\,\sigma_{c,\zeta j}\al{y}
    }}\ ,\nn
\end{eqnarray}
and moreover the interaction terms, derived from the Jaynes-Cummings terms of the previous section, are given by the following expressions. Indeed, in the models ``$Ccq$'' and ``$Scq$'', each qubit corresponding to a cavity of the previous model interacts with two qubits according to the Hamiltonians
\begin{eqnarray}
    H_{c-q,\xi}^\circ&=&
    \hbar\sum_{j=1}^N\frac{g_j}{2}\sum_{\zeta=\pm} \lpg{
    \sigma_{c,\zeta j}\al{x}\pq{\sqrt{\,\ovl n+1}\,\sigma_{\zeta j}\al{x}+\chi_j\,\sqrt{\,\ovl n}\,\sigma_{-\zeta j}\al{x}}
    }\nn\\&&\rpg{
    +
    \sigma_{c,\zeta j}\al{y}\pq{\sqrt{\,\ovl n+1}\,\sigma_{\zeta j}\al{y}-\chi_j\,\sqrt{\,\ovl n}\,\sigma_{-\zeta j}\al{y}}
    }\ ,
    \hspace{0.5cm}{\rm for}\ \xi\in\pg{Ccq,Scq}\ .
    \nn
\end{eqnarray}
Instead, in the models ``$Cq$'' and ``$Sq$'', the Jaynes-Cummings terms result in the anistrotopic $XY$-interaction Hamiltonians
\begin{eqnarray}
    H_{c-q,Cq}^\circ&=&\hbar\sum_{\zeta=\pm} \pq{
    g_1\al{x}\ \sigma_{c,0}\al{x}\ \sigma_{\zeta 1}\al{x}+g_1\al{y}\ \sigma_{c,0}\al{y}\ \sigma_{\zeta 1}\al{y}
    }\ ,
    \nn\\
    H_{c-q,Sq}^\circ&=&\hbar\sum_j\sum_{\zeta=\pm} \pq{
    g_j\al{x}\ \sigma_{c,0}\al{x}\ \sigma_{\zeta j}\al{x}+g_j\al{y}\ \sigma_{c,0}\al{y}\ \sigma_{\zeta j}\al{y}
    }\ ,
    \nn\\&&
    \hspace{4.5cm}
    \nn
\end{eqnarray}
with 
\begin{eqnarray}
    g_j\al{x}&=&\frac{g_j}{2}\ppt{\sqrt{\,\ovl n+1}+\chi_j\,\sqrt{\,\ovl n}}\ ,
    \nn\\
    g_j\al{y}&=&\frac{g_j}{2}\ppt{\sqrt{\,\ovl n+1}-\chi_j\,\sqrt{\,\ovl n}}\ ,
\end{eqnarray}
where $\chi_j$ is defined in Eq.~\rp{chij}.

Now it is easy to check that, as in the previous section, a steady state for these models is
\begin{eqnarray}
    \ke{\Psi^\circ}=\ke{-}_c\ \ke{\psi}\ ,
\end{eqnarray}
where $\ke{-}_c$ indicates the state for all the qubits with lowering operator $\sigma_{c,j}$, where each qubit is in the eigenstate of $\sigma_{c,j}\al{z}$ with eigenvalue $-1$.

\section{Numerical results} \label{num}
The fact that the steady state Eq.~\rp{Psi} is in fact unique, for specific choices of the parameters $\Delta_{c,j}$, $\eta_{c,j}$, $\Delta_{q,j}$, $\eta_{q,j}$, and $g_j$, can be numerically verified. In this section we report the numerical results evaluated in the squeezed representation for various sizes of the arrays of the four models of Sec.~\ref{Sec.c+q}, and by truncating the Hilbert space of the cavities at various Fock numbers. We also include the limiting case of only two levels for the cavity modes, which hence corresponds to the qubits models of Sec.~\ref{Sec.q}.

\begin{figure*}[t!]
    \centering
    \includegraphics[width=16cm]{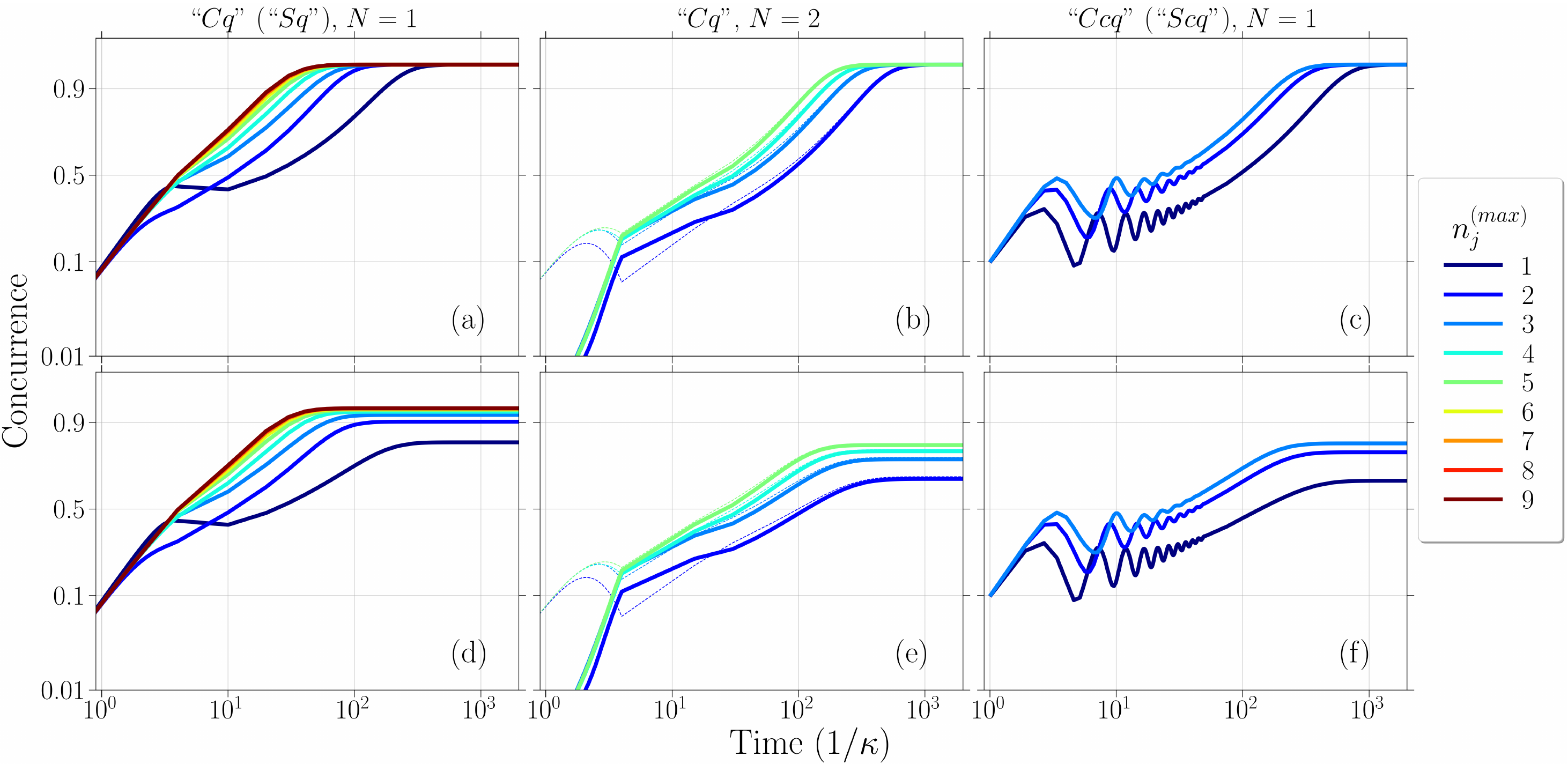}
    \caption{Time evolution of the concurrence for the models: $``Cq"$ with a single qubit pair, $N=1$, in (a) and (d), $``Cq"$ with $N=2$ in (b) and (e), and $``Ccq"$ with $N=1$ in (c) and (f), and for various values of the dimension of the working Hilbert space of the cavities (in the squeezed representation) as reported in the legend. In all the plots $\ovl n=1$. In (b) and (e) the solid (dashed) lines refer to the couple $j=2\,(1)$. In the first row [(a), (b), and (c)] the dephasing rate is $\gamma=0$. In the second row [(d), (e), and (f)] $\gamma=5\times 10^{-4}\kappa$. In (a) and (d) $\Delta_{q,1}\simeq-0.193\kappa$ and $g_1=0.36\kappa$. In (b) and (e) $\Delta_{q,1}\simeq-0.193\kappa$, $\Delta_{q,2}=\Delta_{q,1}+0.05\kappa$, $g_1=0.36\kappa$, and $\eta_{q,1}=0.362\kappa$. In (c) and (f), the dimensions of the Hilbert spaces of all the cavities are truncated at the same Fock number $n_0\al{\rm max}=n_{\pm1}\al{\rm max}$, and $\Delta_{c,1}\simeq-0.26\kappa$, $g_1=0.36\kappa$. 
    The values of the detunings are chosen in order to maximize the decay rate of the arrays for the given choice of interaction strengths (which here are chosen for simplicity of the same order of magnitude). This is done by maximizing the real part of the smallest non-zero value of the total Liouvillian $\LL$ as a function of $\Delta$, see Fig.~\ref{fig4} (c) and Figs.~\ref{fig4.1} (c), (f), and (d) for specific examples.}
    \label{fign}
\end{figure*}

\begin{figure*}[t!]
    \centering
    \includegraphics[width=14cm]{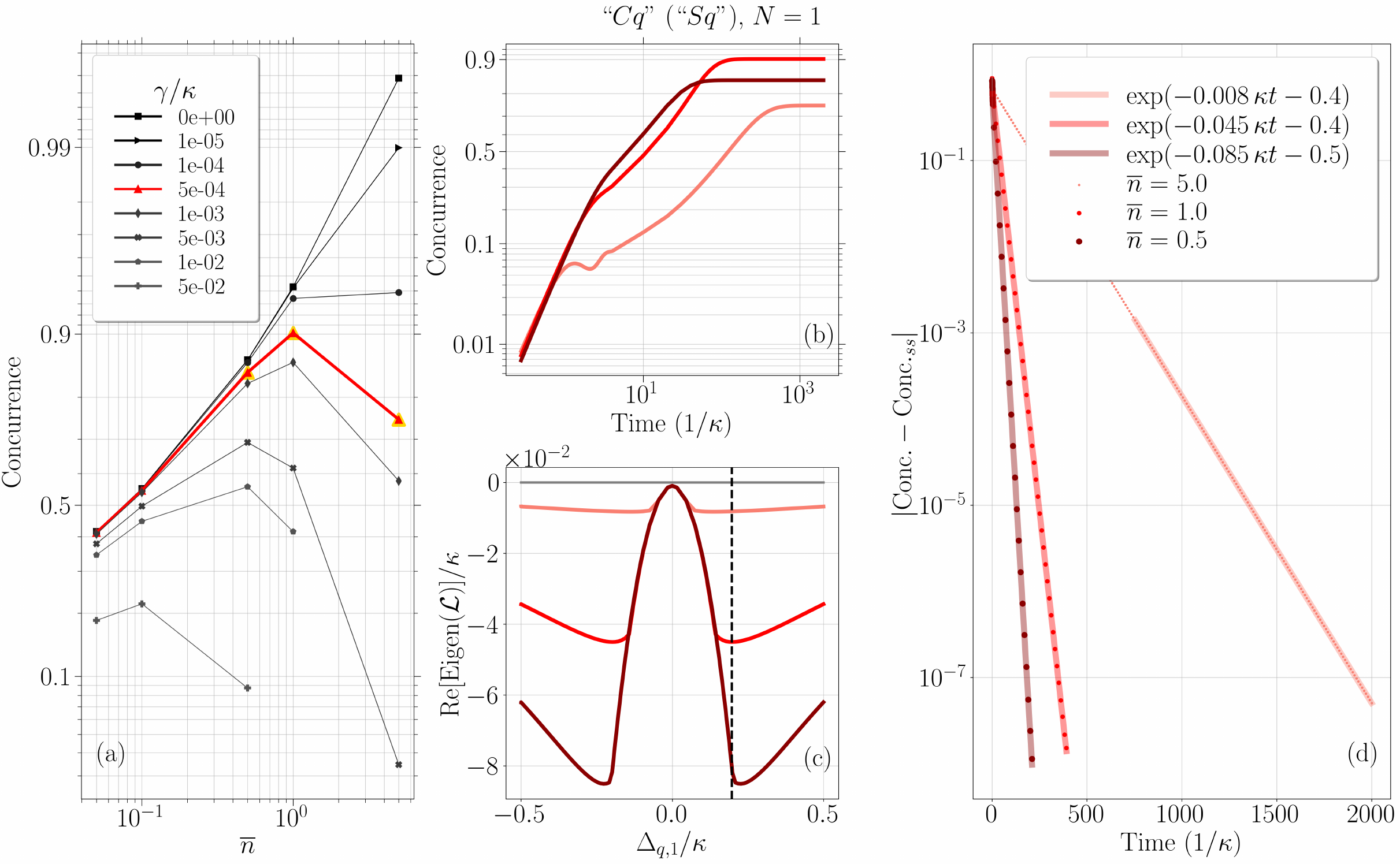}
    \caption{(a) Steady state concurrence for the model $``Cq"$ with $N=1$ (that is equal to $``Sq"$ with $N=1$) as a function of the average number of excitations of the squeezed reservoir $\ovl n$, and for various values of the dephasing rate $\gamma$. (b) Time evolution of the concurrence for the highlighted (yellow) points in (a) corresponding to $\ovl n=0.5$ (dark red), $1$ (red), and $5$ (salmon) with $\gamma=5\times10^{-4}\kappa$. (c) Real part of the first two eigenvalues of $\mathcal L$ for the same points. The initial state is vacuum (in the squeezed representation) for the cavity and the eigenstate of $\sigma_j\al{z}$ with eigenvalue $-1$ for all the qubits. The vertical black, dashed line in (c) indicates the value of $\Delta_{q,1}\simeq0.197\kappa$. The coupling strength is $g_1=0.36\kappa$. The cavity Fock space is truncated at $n_0\al{\rm max}=2$ (in the squeezed representation) to approximate a hybrid model of cavities and qubits. We have verified that similar results hold also for models of only qubits ($n_0\al{\rm max}=1$) and for larger values of $n_0\al{\rm max}$%
    . (d) Decay rate analysis: same results of (b) where we subtract - at each curve - its steady state value, take the modulus,  and fit the result with an exponential decay. The slope of these lines indicate the rate of decay towards the steady state.
    These rates are consistent withe the real part of the eigenvalues of $\LL$ identified by the vertical dashed line in plot (c).
    }
    \label{fig4}
\end{figure*}

\begin{figure*}[t!]
    \includegraphics[width=16cm]{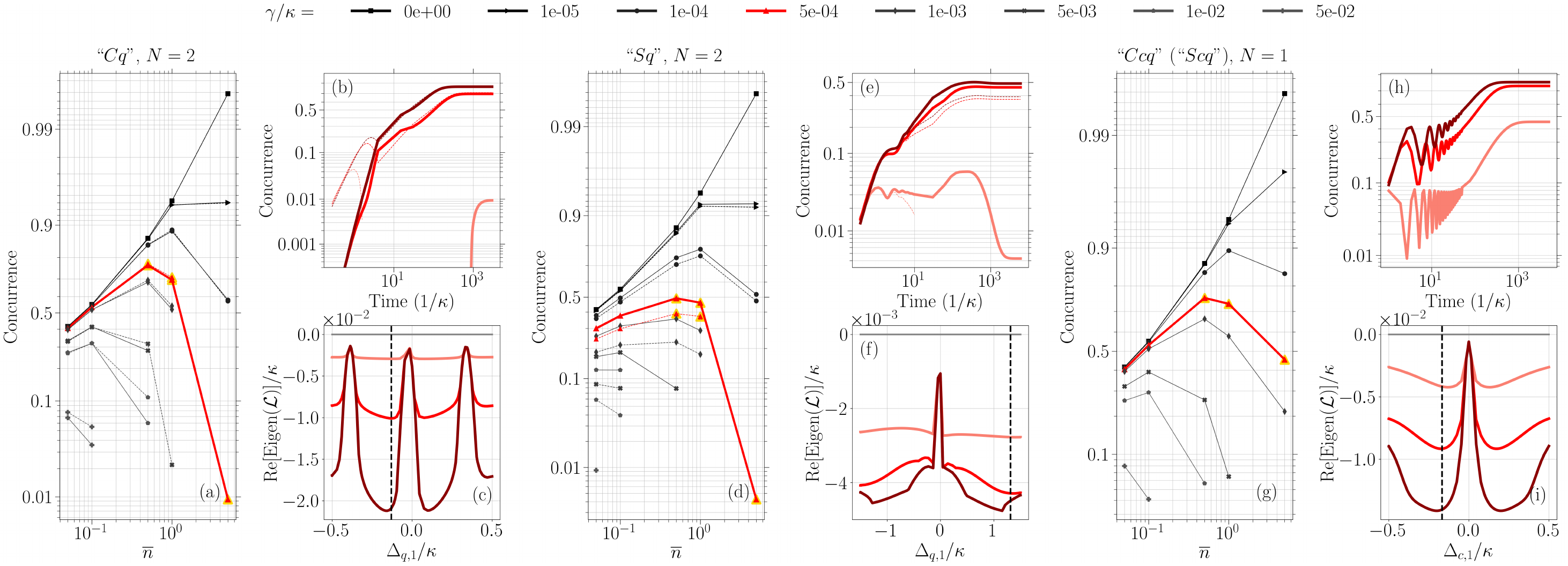}
    \caption{Results similar to those of Fig.~\rp{fig4} (a)-(c) for the models ``$Cq$'' with $N=2$ (a)-(c), ``$Sq$'' with $N=2$ (d)-(f), and ``$Ccq$'' with $N=1$ (that is equal to ``$Scq$" with $N=1$) (g)-(i). In (a) and (d) the solid (dashed) lines refer to the couple $j=2\,(1)$. In (a)-(c) $\Delta_{q,1}\simeq-0.13\kappa$, $\Delta_{q,2}=\Delta_{q,1}+0.05\kappa$, $g_1=0.36\kappa$, and $\eta_{q,1}=0.362\kappa$.
    The Fock space of the cavity is truncated at $n_0\al{\rm max}=2$. In (d)-(f) $\Delta_{q,1}\simeq1.3\kappa$, $\Delta_{q,2}=\Delta_{q,1}+0.05\kappa$, $g_1=0.36\kappa$, and $g_2=0.362\kappa$. The Fock space of the cavity is truncated at $n_0\al{\rm max}=2$. In (g)-(i) $g_1=0.36\kappa$ and $\Delta_{q,1}\simeq -0.17\kappa$. The Fock spaces of the cavities are truncated at $n_0\al{\rm max}=2$ and $n_{\pm1}\al{\rm max}=1$. 
    }
    \label{fig4.1}
\end{figure*}

\begin{figure}[t!]
    \centering
    \includegraphics[width=0.7\textwidth]{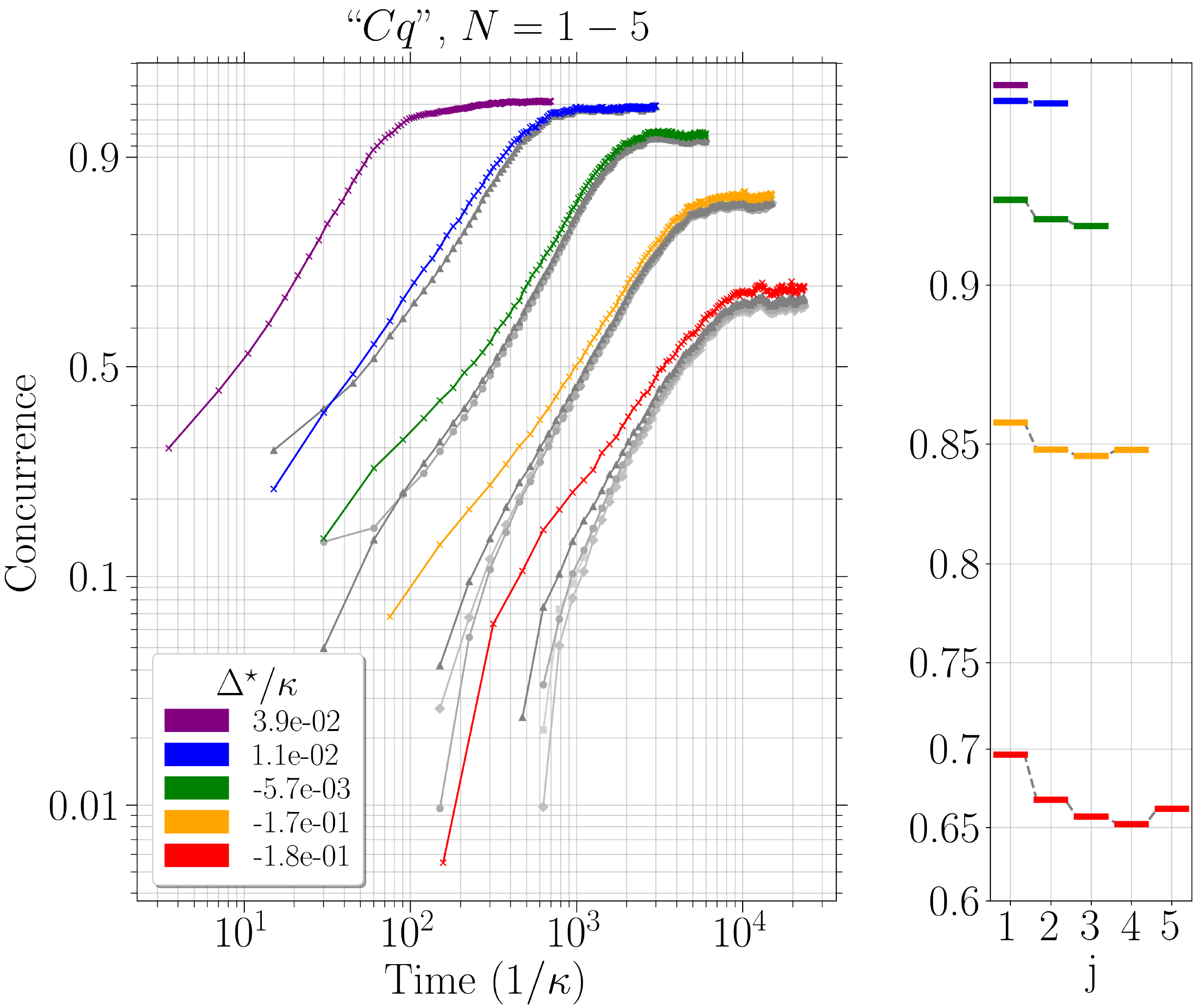}
    \caption{(a) Time evolution of the concurrence of the $j$th couple for different numbers of qubit pairs $N\leq 5$, for ``$Cq$". $N=1$ purple, $N=2$ blue, $N=3$ green, $N=4$ orange, and $N=5$ red. $\overline{n}=1$ and $\gamma/\kappa=10^{-5}$. The different qubit pairs are denoted with different symbols. Cross: $|j|=1$ pair; triangle: $|j|=2$ pair; circle: $|j|=3$ pair; diamond: $|j|=4$ pair; square: $|j|=5$ pair. Only the results for the first pair is suitably coloured, while all the others are gray-scale. The Fock space of the central cavity is truncated to $n_0\al{\rm max}=2$. The system parameters are $g_1=0.36\kappa$, $\eta_{q,j}=g_1+0.002\kappa(j-1)$, and $\Delta_{q,j}=\Delta^\star+0.05\kappa(j-1)$, where the value of $\Delta^\star$ is different for each $N$ as
    reported in the legend, and it has been chosen in order to maximize the concurrence. (b) Corresponding steady state concurrence for all the pairs.
    }
    \label{fig12}
\end{figure}

Additionally, we investigate the sensitivity of these dynamics to extra noise, as depicted in Figs.~\ref{fign}-\ref{fig12}, and examine their scaling behavior with the size of the array, as shown in Fig.~\ref{fig12}. We consider Eq.~\rp{tildeMeq} and include phase noise on the qubits~\footnote{The effect of additional cavity decay on similar systems (made only of cavities) has been analyzed in detail in Refs.~\cite{zippilli2015,zippilli2021}, showing that the entanglement dynamics is unaffected as far as the total additional decay rate is smaller than the coupling rate to the squeezed reservoir $\kappa$.},
according to the equation $\dot{\wt\rho}{}=\LL\ \wt\rho$, where the total Liouvillian superoperator is given by
\begin{eqnarray}\label{LL}
    \LL\ \wt\rho=-\frac{\ii}{\hbar}\pq{\wt H_\xi,\wt \rho}+\frac{\kappa}{2}\,\wt\LL_c\,\wt\rho+\gamma\,\DD\,\wt\rho\ ,
\end{eqnarray}
and where the additional noise on the qubits is described by the Lindblad term 
\begin{eqnarray}
    \DD\ \wt\rho=\sum_{
    \mbox{\scriptsize\ensuremath{\mmat{c}
    {j=-N \\ j\neq 0}
    }}
    }^N\ \sigma_j\al{z}\ \wt\rho\ \sigma_j\al{z}-\wt\rho\ .
\end{eqnarray}%
We highlight that the structure of the Hamiltonians discussed in the previous sections is not sufficient to guarantee that the steady state is unique. A trivial example is when, in the model ``$Sq$'', all the detunings and all the couplings are equal, such that $\Delta_{q,j}=\Delta_{q,j'}$ and $g_j=g_{j'}$ for all $j,j'\in\pg{1,\dots,\,N}$. In this highly symmetric case, any partition of the qubits in pairs may be used to construct a state as that of Eq.~\rp{psiq} which is stationary. All these possible partitions will give rise to a stationary subspace of the system dynamics. However, the actual steady state will depend on the initial state and typically be a statistical mixture of states in this subspace. Therefore, to obtain a unique steady state, such as the pure steady state discussed in Sec.~\ref{Sec:stst} (or an approximate mixed-state version thereof in the presence of finite $\gamma$), it is necessary to avoid these highly symmetric situations, e.g, by using different values of detunings and couplings for each pair.

Here, we characterize the steady state in terms of the concurrence~\cite{horodecki2009} between pairs of qubits. In particular, we verify numerically that, for the chosen set of parameters, only qubits with opposite indices $j$ and $-j$ can get entangled in the steady state. 

We compute the evolution of the system numerically with wave function Monte Carlo techniques, and analyze the stability of the result by truncating the Hilbert space of the cavities to various Fock numbers $n_j\al{\rm max}$ (see Fig.~\ref{fign}).
To ensure the uniqueness of the steady state of the simulated models, we perform numerical analysis on the spectrum of the total Liouvillian. Our results confirm that, for the chosen parameters, the steady state is indeed unique, as evidenced by the null space of the Liouvillian having dimension one. Due to the numerical complexity of the problem, we only consider small arrays. In the squeezed representation one can use a relatively low number of Fock states, with the lowest value $n_j\al{\rm max}=1$ corresponding to models with only qubits of Sec.~\ref{Sec.q} and the largest values of $n_j\al{\rm max}$ which approach the models which include the cavities of Sec.~\ref{Sec.c+q}.
In the case of qubit-only models, we can simulate the full Hilbert space, which allows us to unambiguously demonstrate the uniqueness of the steady state for the chosen parameters. However, for models involving cavities, one can only simulate a finite part of the infinite-dimensional Hilbert space, raising concerns about the effective uniqueness of the steady state for the complete models. Nevertheless, an analysis of Fig.~\ref{fig2} suggests that there are no subspaces outside of the simulated part that do not dissipate and are disconnected from it. This indicates that the steady state is likely to be unique for these cases as well.

In Fig.~\ref{fign} (a)-(c) we observe that in the ideal case ($\gamma=0$) the steady state concurrence is the same for all the models, and independent from the dimension of the Hilbert space. This confirms the result of Eqs.~\rp{psiq} and~\rp{Psi}, i.e., that without any additional dissipation channel the steady state of the qubits is the same for all the models and it depends only on the amount of squeezing of the reservoir which is determined by $\bar n$. However, we notice that the dynamics involving cavities [corresponding to larger values of $n_j\al{\rm max}$] is significantly faster than that of the models with qubits only [$n_j\al{\rm max}=1$]. On the contrary, dephasing reduces the final entanglement, and maximum reduction is observed for the slowest models, namely the models with only qubits [see Figs.~\ref{fign} (d)-(f)].

Figs.~\ref{fig4} and \ref{fig4.1} show that, on the one hand, at zero additional noise ($\gamma=0$) maximum entanglement is achieved for larger values of $\ovl n$, as expected from Eq.~\rp{psiq}, which approaches the product of many Bell, maximally entangled states. On the other hand, when the rate $\gamma$ of the additional noise is finite, maximum entanglement is achieved at \emph{finite} values of $\ovl n$. In fact, the larger $\ovl n$, the slower the dynamics, as illustrated by these figures [see the decay rates for different values of $\bar n$ in Fig.~\ref{fig4} (d)] and, as a consequence, if $\gamma$ is too large noise and decoherence have enough time to spoil the generation of the steady state.

The fact that a larger $\bar n$ corresponds to a slower dynamics is described by Figs.~\ref{fig4} (b), (c), and (d) (and the corresponding plots in Fig.~\ref{fig4.1}). Specifically, in Fig.~\ref{fig4} (c) [and Figs.~\ref{fig4.1} (c), (f), and (i)] we report the real part of the eigenvalues of the total Liouvillian $\LL$ [see Eq.~\rp{LL}]. The smallest (in modulus) real part determines the rate of decay towards the steady state: a larger (in modulus) real part corresponds to a faster dynamics. This is clearly described by Fig.~\ref{fig4} (d), where we report the time evolution of the concurrence relative to its steady state value. This quantity describes an exponential decay with rate given by the values of the eigenvalues identified in Fig.~\ref{fig4} (c). We verified this behavior also for the other models.

In Fig.~\ref{fig12} we verify numerically that the steady state remains unique even for larger number of qubit pairs, when the values of the detuning and of the couplings are properly selected. We show the concurrence for the model ``$Cq$'' with up to $10$ qubits. Both the detuning and the couplings vary linearly with the pair index according, respectively, to the relations $\Delta_{q,j}=\Delta^\star+0.05\kappa(j-1)$ and $\eta_{q,j}=g_1+0.002\kappa(j-1)$. 

In this figure, we also maximize the final concurrence as a function of $\Delta^\star$ only. Specifically, the value of $\Delta^\star$ is chosen in order to maximize the smallest real part of the eigenvalues of $\LL$ for each $N$ (when all the other parameters are kept fixed), which -- as discussed above -- determines how fast the system approaches the steady state. These results show that the final concurrence decreases with the number of pairs and display a behaviour very similar to the entanglement obtained in chains of bosonic modes~\cite{zippilli2015}. 

\begin{figure}
    \centering
    \includegraphics[width=\textwidth]{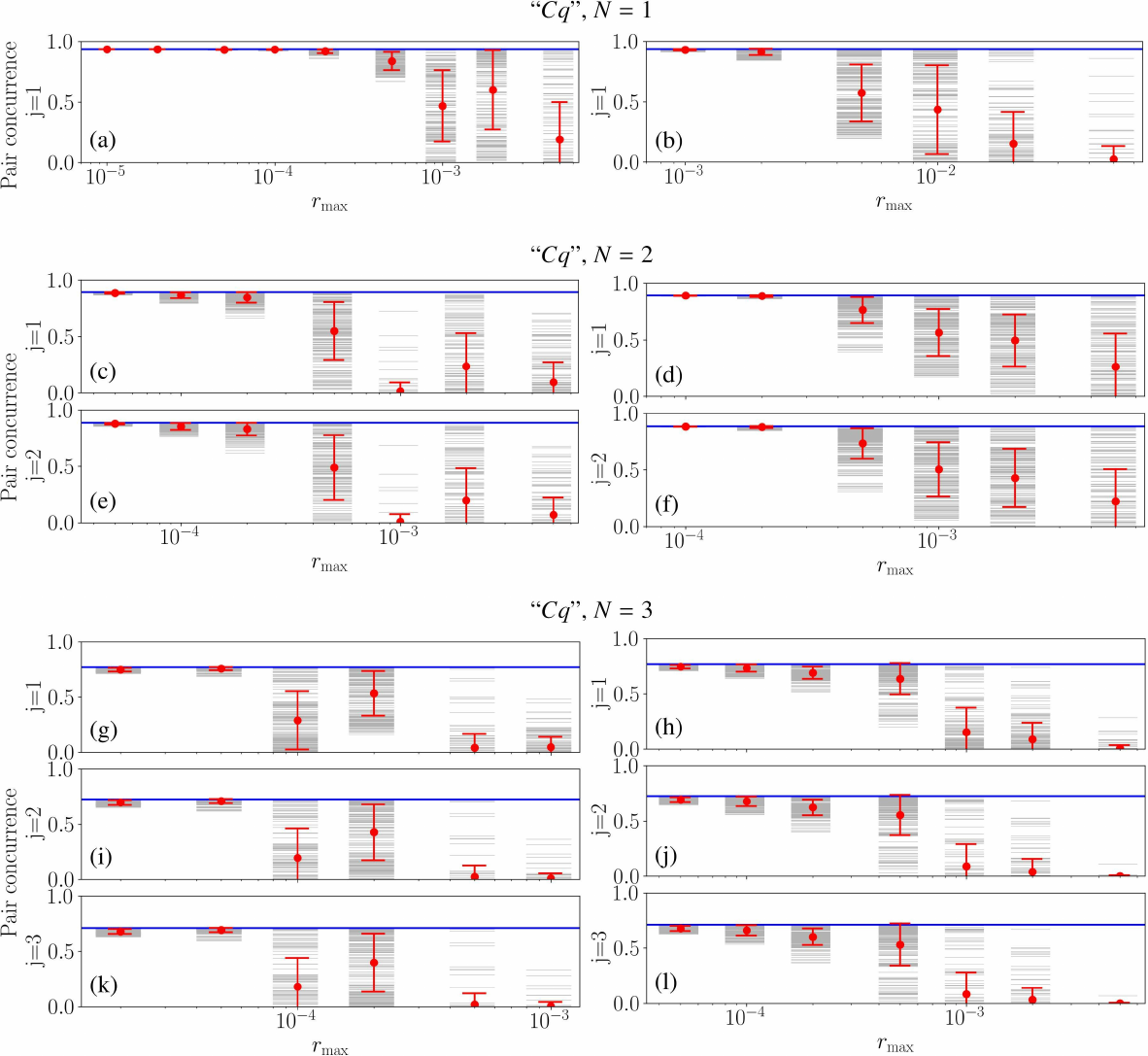}
    \caption{
        Steady state concurrence for the model ``$Cq$'' of only qubits with $N=1,\,2$, and $3$ entangled pairs, evaluated for random variations of the system parameters from the symmetric configuration used in the previous figures.
        The results in the left column are evaluated relaxing the condition of opposite detunings for qubits with opposite indices: in place of $\Delta_{q,j}$ we use $\Delta_{q,\pm j}\al{r}=\pm\Delta_{q,j}\ppt{1+r_{\pm j}\al{\Delta}}$ with $r_{\pm j}\al{\Delta}$ random variables uniformly distributed in the range $\pq{-r_{\rm max},r_{\rm max}}$.
        The results in the right column are evaluated relaxing the condition of equal couplings for qubits with opposite indices: in place of $g_1$ and $\eta_j$ we use $g_{\pm 1}\al{r}=g_1\ppt{1+r_{\pm j}\al{g}}$ and $\eta_{\pm j}\al{r}=\eta_j\ppt{1+r_{\pm j}\al{\eta}}$, respectively, with $r_{\pm 1}\al{g}$ and $r_{\pm j}\al{\eta}$ random variables uniformly distributed in the range $\pq{-r_{\rm max},r_{\rm max}}$.
        In each plot, for each value of $r_{\rm max}$, we report $200$ random realizations (gray lines), the corresponding averages (red dots), and standard deviations (red lines). The blue lines are the ideal results evaluated for the symmetric configuration with $r_j\al{\Delta}=r_j\al{g}=r_j\al{\eta}=0$.
        In all plots $\Delta_j=\ppt{1.8+0.2\,j}\kappa$, $g_1=0.3\,\kappa$, $\eta_j=\ppt{0.45+0.05\,j}\kappa$, and $\gamma=10^{-5}\kappa$.
    }
    \label{fig8}
\end{figure}

\begin{figure}
    \centering
    \includegraphics[width=\textwidth]{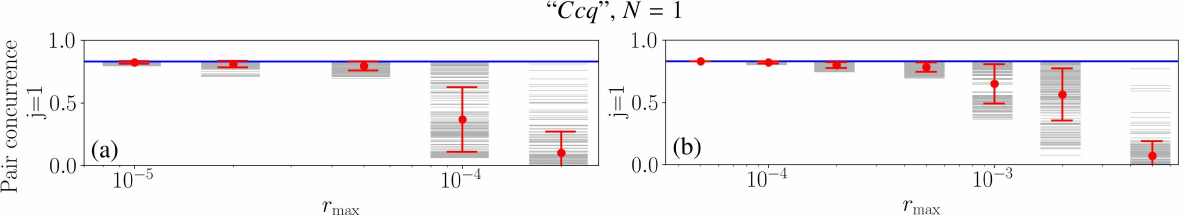}
    \caption{
        As in Fig.~\ref{fig8} (a) and (b) for the model ``$Ccq$'' with only qubits.
    }
    \label{fig9}
\end{figure}
Finally, we analyze, in Figs.~\ref{fig8} and \ref{fig9}, the effect of random deviations of the system parameters from the symmetric configurations  identified in Secs.~\ref{Sec.c+q} and~\ref{Sec.q}. The results show that, as expected, the steady state entanglement tends to be reduced when the system is not symmetric. The reduction is more pronounced for deviation in the values of the detunings than of the couplings, and increases with the system size.

\section{Conclusions}\label{concl}
We have shown that it is possible to drive a quantum array into a pure steady state featuring many entangled qubit pairs, by controlling the dissipative dynamics of a single, central, element, either a cavity or a qubit. The steady state is pure when dissipation acts only on the central element. When additional decoherence is added on the other qubits, the stationary state is an entangled mixed state, and its entanglement is large and robust provided that the decay rate of the additional dephasing processes is much smaller than that of the dissipative decay rate of the central element. 

These models can be realized in a number of different physical situations with atomic systems
~\cite{zhang2017a,brydges2019,kokail2019,
bruzewicz2019,tomza2019}
and solid state nano-devices
~\cite{hacohen-gourgy2015,fitzpatrick2017,roushan2017,
kollar2019,ma2019a,smith2020,kim2021,
lehur2016,gu2017,wendin2017,ozawa2019,carusotto2020,wilkinson2020}, allowing the realization of the chain or of the star geometry.

An interesting related question is whether these approaches can be extended to the preparation of more complex multipartite entangled qubit states, such as the graph states which are the fundamental resource of measurement-based quantum computers~\cite{hein2006}, in a way similar to what has been demonstrated for bosonic modes in Ref.~\cite{zippilli2021}.

\ack 
We acknowledge the support of PNRR MUR project PE0000023-NQSTI (Italy), and of the European Union Horizon 2020 Programme for Research and Innovation through the Project No. 862644 (FET Open QUARTET).

\

\


\begin{thebibliography}{10}

\bibitem{plenio1999}
  
M.~B. Plenio, S.~F. Huelga, A.~Beige, and P.~L. Knight. 
\newblock Cavity-loss-induced generation of entangled atoms 
\newblock{\em Phys. Rev. A}, 59:2468, March 1999.
\newblock \href{https://doi.org/10.1103/PhysRevA.59.2468}{\path{doi:10.1103/PhysRevA.59.2468}}.


\bibitem{kraus2008}
B.~Kraus, H.~P. B{\"u}chler, S.~Diehl, A.~Kantian, A.~Micheli, and P.~Zoller.
\newblock Preparation of entangled states by quantum {{Markov}} processes.
\newblock {\em Phys. Rev. A}, 78(4):042307, October 2008.
\newblock \href {https://doi.org/10.1103/PhysRevA.78.042307}
  {\path{doi:10.1103/PhysRevA.78.042307}}.

\bibitem{diehl2008}
S.~Diehl, A.~Micheli, A.~Kantian, B.~Kraus, H.~P. B{\"u}chler, and P.~Zoller.
\newblock Quantum {{States}} and {{Phases}} in {{Driven Open Quantum Systems}}
  with {{Cold Atoms}}.
\newblock {\em Nat. Phys.}, 4(11):878--883, November 2008.
\newblock \href {https://doi.org/10.1038/nphys1073}
  {\path{doi:10.1038/nphys1073}}.

\bibitem{verstraete2009}
F.~Verstraete, M.~M. Wolf, and J.~I. Cirac.
\newblock Quantum computation and quantum-state engineering driven by
  dissipation.
\newblock {\em Nat. Phys.}, 5(9):633--636, July 2009.
\newblock \href {https://doi.org/10.1038/nphys1342}
  {\path{doi:10.1038/nphys1342}}.

\bibitem{weimer2010}
H.~Weimer, M.~M{\"u}ller, I.~Lesanovsky, P.~Zoller, and H.~P. B{\"u}chler.
\newblock A {{Rydberg}} quantum simulator.
\newblock {\em Nat. Phys.}, 6(5):382--388, May 2010.
\newblock \href {https://doi.org/10.1038/nphys1614}
  {\path{doi:10.1038/nphys1614}}.

\bibitem{barreiro2011}
J.~T. Barreiro, M.~M{\"u}ller, P.~Schindler, D.~Nigg, T.~Monz, M.~Chwalla,
  M.~Hennrich, C.~F. Roos, P.~Zoller, and R.~Blatt.
\newblock An open-system quantum simulator with trapped ions.
\newblock {\em Nature}, 470(7335):486--491, February 2011.
\newblock \href {https://doi.org/10.1038/nature09801}
  {\path{doi:10.1038/nature09801}}.

\bibitem{cho2011a}
J.~Cho, S.~Bose, and M.~S. Kim.
\newblock Optical {{Pumping}} into {{Many-Body Entanglement}}.
\newblock {\em Phys. Rev. Lett.}, 106(2):020504, January 2011.
\newblock \href {https://doi.org/10.1103/PhysRevLett.106.020504}
  {\path{doi:10.1103/PhysRevLett.106.020504}}.

\bibitem{koga2012}
K.~Koga and N.~Yamamoto.
\newblock Dissipation-induced pure {{Gaussian}} state.
\newblock {\em Phys. Rev. A}, 85(2):022103, February 2012.
\newblock \href {https://doi.org/10.1103/PhysRevA.85.022103}
  {\path{doi:10.1103/PhysRevA.85.022103}}.

\bibitem{morigi2015a}
G.~Morigi, J.~Eschner, C.~Cormick, Y.~Lin, D.~Leibfried, and D.~J. Wineland.
\newblock Dissipative {{Quantum Control}} of a {{Spin Chain}}.
\newblock {\em Phys. Rev. Lett.}, 115(20):200502, November 2015.
\newblock \href {https://doi.org/10.1103/PhysRevLett.115.200502}
  {\path{doi:10.1103/PhysRevLett.115.200502}}.

\bibitem{reiter2016}
F.~Reiter, D.~Reeb, and A.~S. S{\o}rensen.
\newblock Scalable {{Dissipative Preparation}} of {{Many-Body Entanglement}}.
\newblock {\em Phys. Rev. Lett.}, 117(4):040501, July 2016.
\newblock \href {https://doi.org/10.1103/PhysRevLett.117.040501}
  {\path{doi:10.1103/PhysRevLett.117.040501}}.

\bibitem{zippilli2013_zippilli2013a}
S.~Zippilli, M.~Paternostro, G.~Adesso, and F.~Illuminati.
\newblock Entanglement {{Replication}} in {{Driven Dissipative Many-Body}}
  systems.
\newblock {\em Phys. Rev. Lett.}, 110(4):040503, January 2013.
\newblock \href {https://doi.org/10.1103/PhysRevLett.110.040503}
  {\path{doi:10.1103/PhysRevLett.110.040503}};
\newblock Erratum: {{Entanglement Replication}} in {{Driven Dissipative
  Many-Body Systems}} [{{Phys}}. {{Rev}}. {{Lett}}. 110, 040503 (2013)].
\newblock {\em Phys. Rev. Lett.}, 111(16):169901, October 2013.
\newblock \href {https://doi.org/10.1103/PhysRevLett.111.169901}
  {\path{doi:10.1103/PhysRevLett.111.169901}}.

\bibitem{zippilli2014}
S.~Zippilli and F.~Illuminati.
\newblock Non-{{Markovian}} dynamics and steady-state entanglement of cavity
  arrays in finite-bandwidth squeezed reservoirs.
\newblock {\em Phys. Rev. A}, 89(3):033803, March 2014.
\newblock \href {https://doi.org/10.1103/PhysRevA.89.033803}
  {\path{doi:10.1103/PhysRevA.89.033803}}.

\bibitem{ma2019}
S.~Ma and M.~J. Woolley.
\newblock Entangled pure steady states in harmonic chains with a two-mode
  squeezed reservoir.
\newblock {\em J. Phys. A: Math. Theor.}, 52(32):325301, July 2019.
\newblock \href {https://doi.org/10.1088/1751-8121/ab2ce7}
  {\path{doi:10.1088/1751-8121/ab2ce7}}.

\bibitem{wendenbaum2015}
P.~Wendenbaum, T.~Platini, and D.~Karevski.
\newblock Entanglement replication via quantum repeated interactions.
\newblock {\em Phys. Rev. A}, 91(4):040303, April 2015.
\newblock \href {https://doi.org/10.1103/PhysRevA.91.040303}
  {\path{doi:10.1103/PhysRevA.91.040303}}.

\bibitem{pocklington2022}
A.~Pocklington, Y.-X. Wang, Y.~Yanay, and A.~A. Clerk.
\newblock Stabilizing volume-law entangled states of fermions and qubits using local dissipation.
\newblock {\em Phys. Rev. B}, 105(14):L140301, April 2022.
\newblock \href {https://doi.org/10.1103/PhysRevB.105.L140301}
  {\path{doi:10.1103/PhysRevB.105.L140301}}.

\bibitem{zippilli2015}
S.~Zippilli, J.~Li, and D.~Vitali.
\newblock Steady-state nested entanglement structures in harmonic chains with
  single-site squeezing manipulation.
\newblock {\em Phys. Rev. A}, 92(3):032319, September 2015.
\newblock \href {https://doi.org/10.1103/PhysRevA.92.032319}
  {\path{doi:10.1103/PhysRevA.92.032319}}.

\bibitem{asjad2016a}
M.~Asjad, S.~Zippilli, and D.~Vitali.
\newblock Mechanical {{Einstein-Podolsky-Rosen}} entanglement with a
  finite-bandwidth squeezed reservoir.
\newblock {\em Phys. Rev. A}, 93(6):062307, June 2016.
\newblock \href {https://doi.org/10.1103/PhysRevA.93.062307}
  {\path{doi:10.1103/PhysRevA.93.062307}}.

\bibitem{ma2017}
S.~Ma, M.~J. Woolley, I.~R. Petersen, and N.~Yamamoto.
\newblock Pure {{Gaussian}} states from quantum harmonic oscillator chains with
  a single local dissipative process.
\newblock {\em J. Phys. A: Math. Theor.}, 50(13):135301, 2017.
\newblock \href {https://doi.org/10.1088/1751-8121/aa5fbe}
  {\path{doi:10.1088/1751-8121/aa5fbe}}.

\bibitem{yanay2018}
Y.~Yanay and A.~A. Clerk.
\newblock Reservoir engineering of bosonic lattices using chiral symmetry and
  localized dissipation.
\newblock {\em Phys. Rev. A}, 98(4):043615, October 2018.
\newblock \href {https://doi.org/10.1103/PhysRevA.98.043615}
  {\path{doi:10.1103/PhysRevA.98.043615}}.

\bibitem{yanay2020b}
Y.~Yanay and A.~A. Clerk.
\newblock Reservoir engineering with localized dissipation: {{Dynamics}} and
  prethermalization.
\newblock {\em Phys. Rev. Research}, 2(2):023177, May 2020.
\newblock \href {https://doi.org/10.1103/PhysRevResearch.2.023177}
  {\path{doi:10.1103/PhysRevResearch.2.023177}}.

\bibitem{yanay2020a}
Y.~Yanay.
\newblock Algorithm for tailoring a quadratic lattice with a local squeezed
  reservoir to stabilize generic chiral states with nonlocal entanglement.
\newblock {\em Phys. Rev. A}, 102(3):032417, September 2020.
\newblock \href {https://doi.org/10.1103/PhysRevA.102.032417}
  {\path{doi:10.1103/PhysRevA.102.032417}}.

\bibitem{zippilli2021}
S.~Zippilli and D.~Vitali.
\newblock Dissipative {{Engineering}} of {{Gaussian Entangled States}} in
  {{Harmonic Lattices}} with a {{Single-Site Squeezed Reservoir}}.
\newblock {\em Phys. Rev. Lett.}, 126(2):020402, January 2021.
\newblock \href {https://doi.org/10.1103/PhysRevLett.126.020402}
  {\path{doi:10.1103/PhysRevLett.126.020402}}.


\bibitem{ma2021}  
S.~Ma, J.~Zhang, X.~Li, Y.~Ren, J.~Xie, M.~Cao, and F.~Li, 
\newblock Coupling-modulation–mediated generation of stable entanglement of superconducting qubits via dissipation 
\newblock {\em EPL}, 135:63001, 2021.
\href{https://doi.org/10.1209/0295-5075/ac2b5c}{\path{doi:10.1209/0295-5075/ac2b5c}}.


\bibitem{asjad2016b}

M.~Asjad, S.~Zippilli, and D.~Vitali,
\newblock Suppression of Stokes scattering and improved optomechanical cooling with squeezed light
\newblock {\em Phys. Rev. A}, 94(5):051801, November 2016.
\newblock \href {https://link.aps.org/doi/10.1103/PhysRevA.94.051801}
  {\path{doi:10.1103/PhysRevA.94.051801}}.


\bibitem{Clark_2017}

J.~B. Clark, F.~Lecocq, R.~W. Simmonds, J.~Aumentado, and J.~D. Teufel. 
\newblock Sideband cooling beyond the quantum backaction limit with squeezed light
\newblock Nature 541, 191–195 (2017). 
\newblock \href{https://doi.org/10.1038/nature20604}
  {\path{doi:10.1038/nature20604}}.




\bibitem{virgocollaboration2019}  

F.~Acernese, \textit{et al.} (Virgo Collaboration).
Increasing the Astrophysical Reach of the Advanced Virgo Detector via the Application of Squeezed Vacuum States of Light. 
{\em Phys. Rev. Lett.}, 123:231108, 2019.
\href{https://doi.org/10.1103/PhysRevLett.123.231108}
{\path{doi:10.1103/PhysRevLett.123.231108}}.


\bibitem{tse2019}  

M.~Tse, H.~Yu, \textit{et al.}. 
Quantum-Enhanced Advanced LIGO Detectors in the Era of Gravitational-Wave Astronomy. 
{\em Phys. Rev. Lett.}, 123:231107, 2019.
\href{https://doi.org/10.1103/PhysRevLett.123.231107}
{\path{doi:10.1103/PhysRevLett.123.231107}}.



\bibitem{vitagliano2010}
G.~Vitagliano, A.~Riera, and J.~I. Latorre.
\newblock Volume-law scaling for the entanglement entropy in spin-1/2 chains.
\newblock {\em New J. Phys.}, 12(11):113049, November 2010.
\newblock \href {https://doi.org/10.1088/1367-2630/12/11/113049}
  {\path{doi:10.1088/1367-2630/12/11/113049}}.

\bibitem{ramirez2014}
G. Ram{\'i}rez, J. {Rodr{\'i}guez-Laguna}, and G. Sierra.
\newblock From conformal to volume law for the entanglement entropy in
  exponentially deformed critical spin 1/2 chains.
\newblock {\em J. Stat. Mech.}, 2014(10):P10004, October 2014.
\newblock \href {https://doi.org/10.1088/1742-5468/2014/10/P10004}
  {\path{doi:10.1088/1742-5468/2014/10/P10004}}.

\bibitem{langlett2022}
C.~M. Langlett, Z.-C. Yang, J. Wildeboer, A.~V. Gorshkov,
  T. Iadecola, and S. Xu.
\newblock Rainbow scars: {{From}} area to volume law.
\newblock {\em Phys. Rev. B}, 105(6):L060301, February 2022.
\newblock \href {https://doi.org/10.1103/PhysRevB.105.L060301}
  {\path{doi:10.1103/PhysRevB.105.L060301}}.

\bibitem{zippilli2013b}
S.~Zippilli, S.~M. Giampaolo, and F.~Illuminati.
\newblock Surface entanglement in quantum spin networks.
\newblock {\em Phys. Rev. A}, 87(4):042304, April 2013.
\newblock \href {https://doi.org/10.1103/PhysRevA.87.042304}
  {\path{doi:10.1103/PhysRevA.87.042304}}.

\bibitem{difranco2008}
C.~Di~Franco, M.~Paternostro, and M.~S. Kim.
\newblock Nested entangled states for distributed quantum channels.
\newblock {\em Phys. Rev. A}, 77(2), February 2008.
\newblock \href {https://doi.org/10.1103/PhysRevA.77.020303}
  {\path{doi:10.1103/PhysRevA.77.020303}}.

\bibitem{alkurtass2014}
B. Alkurtass, L. Banchi, and S. Bose.
\newblock Optimal quench for distance-independent entanglement and maximal
  block entropy.
\newblock {\em Phys. Rev. A}, 90(4):042304, October 2014.
\newblock \href {https://doi.org/10.1103/PhysRevA.90.042304}
  {\path{doi:10.1103/PhysRevA.90.042304}}.

\bibitem{pitsios2017}
I. Pitsios, L. Banchi, A.~S. Rab, M. Bentivegna, D.
  Caprara, A. Crespi, N. Spagnolo, S. Bose, P. Mataloni,
  R. Osellame, and F. Sciarrino.
\newblock Photonic simulation of entanglement growth and engineering after a
  spin chain quench.
\newblock {\em Nat Commun}, 8(1):1569, November 2017.
\newblock \href {https://doi.org/10.1038/s41467-017-01589-y}
  {\path{doi:10.1038/s41467-017-01589-y}}.

\bibitem{cottrell2019}
W. Cottrell, B. Freivogel, D.~M. Hofman, and S.~F. Lokhande.
\newblock How to build the thermofield double state.
\newblock {\em J. High Energ. Phys.}, 2019(2):58, February 2019.
\newblock \href {https://doi.org/10.1007/JHEP02(2019)058}
  {\path{doi:10.1007/JHEP02(2019)058}}.

\bibitem{brown2021}
A.~R. Brown, H. Gharibyan, S. Leichenauer, H.~W. Lin, S.
  Nezami, G. Salton, L. Susskind, B. Swingle, and M. Walter.
\newblock Quantum {{Gravity}} in the {{Lab}}: {{Teleportation}} by {{Size}} and
  {{Traversable Wormholes}}.
\newblock {\em arXiv:1911.06314}, February 2021.

\bibitem{dutta2020}
S.~Dutta and N.~R. Cooper.
\newblock Long-{{Range Coherence}} and {{Multiple Steady States}} in a {{Lossy
  Qubit Array}}.
\newblock {\em Phys. Rev. Lett.}, 125(24):240404, December 2020.
\newblock \href {https://doi.org/10.1103/PhysRevLett.125.240404}
  {\path{doi:10.1103/PhysRevLett.125.240404}}.

\bibitem{dutta2021}
S.~Dutta and N.~R. Cooper.
\newblock Out-of-equilibrium steady states of a locally driven lossy qubit
  array.
\newblock {\em Phys. Rev. Research}, 3(1):L012016, February 2021.
\newblock \href {https://doi.org/10.1103/PhysRevResearch.3.L012016}
  {\path{doi:10.1103/PhysRevResearch.3.L012016}}.

\bibitem{horodecki2009}
R.~Horodecki, P.~Horodecki, M.~Horodecki, and K.~Horodecki.
\newblock Quantum entanglement.
\newblock {\em Rev. Mod. Phys.}, 81(2):865--942, June 2009.
\newblock \href {https://doi.org/10.1103/RevModPhys.81.865}
  {\path{doi:10.1103/RevModPhys.81.865}}.



\bibitem{LocalGlobal_MEQ}

J. Onam González, L. A. Correa, G. Nocerino, J. P. Palao, D. Alonso, G. Adesso. Testing the Validity of the 'Local' and 'Global' GKLS Master Equations on an Exactly Solvable Model. {\em Open Syst. Inf. Dyn.}, 24:1740010, 2017.
\href{https://doi.org/10.1142/S1230161217400108}{\path{doi:10.1142/S1230161217400108}};

%

P. P. Hofer, M. Perarnau-Llobet, L. David M. Miranda, G. Haack, R. Silva, J. Bohr Brask, N. Brunner. Markovian master equations for quantum thermal machines: local versus global approach.
{\em New J. Phys.}, 19:123037, 2017.
\href{https://doi.org/10.1088/1367-2630/aa964f}{\path{doi:10.1088/1367-2630/aa964f}};

%

M. Cattaneo, G. L. Giorgi, S. Maniscalco, R. Zambrini. Local versus global master equation with common and separate baths: superiority of the global approach in partial secular approximation. 
{\em New J. Phys.}, 21:113045, 2019.
\href{https://doi.org/10.1088/1367-2630/ab54ac}{\path{doi:10.1088/1367-2630/ab54ac}};

%

S. Scali, J. Anders, L. A. Correa. Local master equations bypass the secular approximation. 
{\em Quantum}, 5:451, 2021.
\href{https://doi.org/10.22331/q-2021-05-01-451}{\path{doi:10.22331/q-2021-05-01-451}}.




\bibitem{benatti2003_kraus2004_paternostro2004_adesso2010}
F.~Benatti, R.~Floreanini, and M.~Piani.
\newblock Environment {{Induced Entanglement}} in {{Markovian Dissipative
  Dynamics}}.
\newblock {\em Phys. Rev. Lett.}, 91(7):070402, 2003.
\newblock \href {https://doi.org/10.1103/PhysRevLett.91.070402}
  {\path{doi:10.1103/PhysRevLett.91.070402}}.;
B.~Kraus and J.~I. Cirac.
\newblock Discrete {{Entanglement Distribution}} with {{Squeezed Light}}.
\newblock {\em Phys. Rev. Lett.}, 92(1):013602, January 2004.
\newblock \href {https://doi.org/10.1103/PhysRevLett.92.013602}
  {\path{doi:10.1103/PhysRevLett.92.013602}};
M.~Paternostro, W.~Son, and M.~S. Kim.
\newblock Complete {{Conditions}} for {{Entanglement Transfer}}.
\newblock {\em Phys. Rev. Lett.}, 92(19):197901, May 2004.
\newblock \href {https://doi.org/10.1103/PhysRevLett.92.197901}
  {\path{doi:10.1103/PhysRevLett.92.197901}}.
G.~Adesso, S.~Campbell, F.~Illuminati, and M.~Paternostro.
\newblock Controllable {{Gaussian-Qubit Interface}} for {{Extremal Quantum
  State Engineering}}.
\newblock {\em Phys. Rev. Lett.}, 104(24):240501, June 2010.
\newblock \href {https://doi.org/10.1103/PhysRevLett.104.240501}
  {\path{doi:10.1103/PhysRevLett.104.240501}}.

\bibitem{zhang2017a}
J.~Zhang, G.~Pagano, P.~W. Hess, A.~Kyprianidis, P.~Becker, H.~Kaplan, A.~V.
  Gorshkov, Z.-X. Gong, and C.~Monroe.
\newblock Observation of a {{Many-Body Dynamical Phase Transition}} with a
  53-{{Qubit Quantum Simulator}}.
\newblock {\em Nature}, 551(7682):601--604, November 2017.
\newblock \href {https://doi.org/10.1038/nature24654}
  {\path{doi:10.1038/nature24654}}.

\bibitem{brydges2019}
T.~Brydges, A.~Elben, P.~Jurcevic, B.~Vermersch, C.~Maier, B.~P. Lanyon,
  P.~Zoller, R.~Blatt, and C.~F. Roos.
\newblock Probing {{R\'enyi}} entanglement entropy via randomized measurements.
\newblock {\em Science}, 364(6437):260--263, April 2019.
\newblock \href {https://doi.org/10.1126/science.aau4963}
  {\path{doi:10.1126/science.aau4963}}.

\bibitem{kokail2019}
C.~Kokail, C.~Maier, R.~{van Bijnen}, T.~Brydges, M.~K. Joshi, P.~Jurcevic,
  C.~A. Muschik, P.~Silvi, R.~Blatt, C.~F. Roos, and P.~Zoller.
\newblock Self-verifying variational quantum simulation of lattice models.
\newblock {\em Nature}, 569(7756):355--360, May 2019.
\newblock \href {https://doi.org/10.1038/s41586-019-1177-4}
  {\path{doi:10.1038/s41586-019-1177-4}}.

\bibitem{bruzewicz2019}
C.~D. Bruzewicz, J.~Chiaverini, R.~McConnell, and J.~M. Sage.
\newblock Trapped-{{Ion Quantum Computing}}: {{Progress}} and {{Challenges}}.
\newblock {\em Applied Physics Reviews}, 6(2):021314, June 2019.
\newblock \href {https://doi.org/10.1063/1.5088164}
  {\path{doi:10.1063/1.5088164}}.

\bibitem{tomza2019}
M.~Tomza, K.~Jachymski, R.~Gerritsma, A.~Negretti, T.~Calarco, Z.~Idziaszek,
  and P.~S. Julienne.
\newblock Cold hybrid ion-atom systems.
\newblock {\em Reviews of Modern Physics}, 91:035001, July 2019.
\newblock \href {https://doi.org/10.1103/RevModPhys.91.035001}
  {\path{doi:10.1103/RevModPhys.91.035001}}.

\bibitem{hacohen-gourgy2015}
S.~{Hacohen-Gourgy}, V.~V. Ramasesh, C.~De~Grandi, I.~Siddiqi, and S.~M.
  Girvin.
\newblock Cooling and {{Autonomous Feedback}} in a {{Bose-Hubbard Chain}} with
  {{Attractive Interactions}}.
\newblock {\em Phys. Rev. Lett.}, 115(24):240501, December 2015.
\newblock \href {https://doi.org/10.1103/PhysRevLett.115.240501}
  {\path{doi:10.1103/PhysRevLett.115.240501}}.

\bibitem{fitzpatrick2017}
M.~Fitzpatrick, N.~M. Sundaresan, A.~C.~Y. Li, J.~Koch, and A.~A. Houck.
\newblock Observation of a {{Dissipative Phase Transition}} in a
  {{One-Dimensional Circuit QED Lattice}}.
\newblock {\em Phys. Rev. X}, 7(1):011016, February 2017.
\newblock \href {https://doi.org/10.1103/PhysRevX.7.011016}
  {\path{doi:10.1103/PhysRevX.7.011016}}.

\bibitem{roushan2017}
P.~Roushan, C.~Neill, J.~Tangpanitanon, V.~M. Bastidas, A.~Megrant, R.~Barends,
  Y.~Chen, Z.~Chen, B.~Chiaro, A.~Dunsworth, A.~Fowler, B.~Foxen, M.~Giustina,
  E.~Jeffrey, J.~Kelly, E.~Lucero, J.~Mutus, M.~Neeley, C.~Quintana, D.~Sank,
  A.~Vainsencher, J.~Wenner, T.~White, H.~Neven, D.~G. Angelakis, and
  J.~Martinis.
\newblock Spectroscopic signatures of localization with interacting photons in
  superconducting qubits.
\newblock {\em Science}, 358(6367):1175--1179, December 2017.
\newblock \href {https://doi.org/10.1126/science.aao1401}
  {\path{doi:10.1126/science.aao1401}}.

\bibitem{kollar2019}
A.~J. Koll{\'a}r, M.~Fitzpatrick, and A.~A. Houck.
\newblock Hyperbolic lattices in circuit quantum electrodynamics.
\newblock {\em Nature}, 571(7763):45--50, July 2019.
\newblock \href {https://doi.org/10.1038/s41586-019-1348-3}
  {\path{doi:10.1038/s41586-019-1348-3}}.

\bibitem{ma2019a}
R.~Ma, B.~Saxberg, C.~Owens, N.~Leung, Y.~Lu, J.~Simon, and D.~I. Schuster.
\newblock A dissipatively stabilized {{Mott}} insulator of photons.
\newblock {\em Nature}, 566(7742):51--57, February 2019.
\newblock \href {https://doi.org/10.1038/s41586-019-0897-9}
  {\path{doi:10.1038/s41586-019-0897-9}}.

\bibitem{smith2020}
A.~Smith, B.~Jobst, A.~G. Green, and F.~Pollmann.
\newblock Crossing a topological phase transition with a quantum computer.
\newblock {\em Phys. Rev. Research}, 4(2):L022020, April 2022.
\newblock \href {https://doi.org/10.1103/PhysRevResearch.4.L022020}
  {\path{doi:10.1103/PhysRevResearch.4.L022020}}.

\bibitem{kim2021}
E.~Kim, X.~Zhang, V.~S. Ferreira, J.~Banker, J.~K. Iverson, A.~Sipahigil,
  M.~Bello, A.~{Gonz{\'a}lez-Tudela}, M.~Mirhosseini, and O.~Painter.
\newblock Quantum {{Electrodynamics}} in a {{Topological Waveguide}}.
\newblock {\em Phys. Rev. X}, 11(1):011015, January 2021.
\newblock \href {https://doi.org/10.1103/PhysRevX.11.011015}
  {\path{doi:10.1103/PhysRevX.11.011015}}.

\bibitem{lehur2016}
K.~Le~Hur, L.~Henriet, A.~Petrescu, K.~Plekhanov, G.~Roux, and M.~Schir{\'o}.
\newblock Many-body quantum electrodynamics networks: {{Non-equilibrium}}
  condensed matter physics with light.
\newblock {\em Comptes Rendus Physique}, 17(8):808--835, October 2016.
\newblock \href {https://doi.org/10.1016/j.crhy.2016.05.003}
  {\path{doi:10.1016/j.crhy.2016.05.003}}.

\bibitem{gu2017}
X.~Gu, A.~F. Kockum, A.~Miranowicz, Y.-X. Liu, and F.~Nori.
\newblock Microwave photonics with superconducting quantum circuits.
\newblock {\em Physics Reports}, 718--719:1--102, November 2017.
\newblock \href {https://doi.org/10.1016/j.physrep.2017.10.002}
  {\path{doi:10.1016/j.physrep.2017.10.002}}.

\bibitem{wendin2017}
G.~Wendin.
\newblock Quantum information processing with superconducting circuits: A
  review.
\newblock {\em Rep. Prog. Phys.}, 80(10):106001, 2017.
\newblock \href {https://doi.org/10.1088/1361-6633/aa7e1a}
  {\path{doi:10.1088/1361-6633/aa7e1a}}.

\bibitem{ozawa2019}
T.~Ozawa, H.~M. Price, A.~Amo, N.~Goldman, M.~Hafezi, L.~Lu, M.~C. Rechtsman,
  D.~Schuster, J.~Simon, O.~Zilberberg, and I.~Carusotto.
\newblock Topological photonics.
\newblock {\em Reviews of Modern Physics}, 91:015006, January 2019.
\newblock \href {https://doi.org/10.1103/RevModPhys.91.015006}
  {\path{doi:10.1103/RevModPhys.91.015006}}.

\bibitem{carusotto2020}
I.~Carusotto, A.~A. Houck, A.~J. Koll{\'a}r, P.~Roushan, D.~I. Schuster, and
  J.~Simon.
\newblock Photonic materials in circuit quantum electrodynamics.
\newblock {\em Nat. Phys.}, 16(3):268--279, March 2020.
\newblock \href {https://doi.org/10.1038/s41567-020-0815-y}
  {\path{doi:10.1038/s41567-020-0815-y}}.

\bibitem{wilkinson2020}
S.~A. Wilkinson and M.~J. Hartmann.
\newblock Superconducting quantum many-body circuits for quantum simulation and
  computing.
\newblock {\em Appl. Phys. Lett.}, 116(23):230501, June 2020.
\newblock \href {https://doi.org/10.1063/5.0008202}
  {\path{doi:10.1063/5.0008202}}.

\bibitem{hein2006}
M.~Hein, W.~D{\"u}r, J.~Eisert, R.~Raussendorf, M.~{Van den Nest}, and H.-J.
  Briegel.
\newblock Entanglement in graph states and its applications.
\newblock {\em Proc. Int. Sch. Phys. Enrico Fermi}, 162(Quantum Computers,
  Algorithms and Chaos):115--218, 2006.
\newblock \href {https://doi.org/10.3254/978-1-61499-018-5-115}
  {\path{doi:10.3254/978-1-61499-018-5-115}}.

\end{thebibliography}
\end{document}